\documentclass[11pt]{article}
\usepackage[dvips]{graphicx}
\usepackage{amsfonts}
\usepackage{amssymb}
\usepackage{amsmath}
\usepackage{epsfig}
\usepackage{cite}
\usepackage{multirow}
\usepackage{latexsym}
\usepackage{graphicx}
\usepackage{float}

\newcommand{\nn}{\nonumber}
\newcommand{\be}{\begin{equation}}
\newcommand{\ee}{\end{equation}}
\newcommand{\bea}{\begin{eqnarray}}
\newcommand{\eea}{\end{eqnarray}}

\newcommand{\beqa}{\begin{eqnarray}}
\newcommand{\eeqa}{\end{eqnarray}}

\newcommand{\un}{{\bf 1}}

\newcommand{\f}{{\bf 5}}
\newcommand{\fb}{{\bf \bar{5}}}
\newcommand{\te}{{\bf 10}}
\newcommand{\teb}{{\bf \bar{10}}}
\newcommand{\op}{\oplus}

\newcommand{\ord}{\mathrm{ord}}

\topmargin
-1.5cm
\textwidth
15.5cm
\textheight
23.5cm
\oddsidemargin
0.7cm
\evensidemargin
1.2cm

\begin{document}

\makeatletter
\@addtoreset{equation}{section}
\makeatother
\renewcommand{\theequation}{\thesection.\arabic{equation}}
\pagestyle{empty}
\rightline{CPHT-RR050.0610, LPT-ORSAY 10-51}
\vspace{1.9cm}
\begin{center}
\Large{\bf On hypercharge flux and exotics in F-theory GUTs \\[12mm]}
\large{Emilian Dudas$^{1,2}$,\; Eran Palti$^1$ \\[5mm]}
\small{$^1$ Centre de Physique Th´eorique, Ecole Polytechnique, CNRS, 91128 Palaiseau, France.}\\
\small{$^2$ LPT, UMR du CNRS 8627, Bˆat 210, Universit´e de Paris-Sud, 91405 Orsay Cedex, France.} \\[3mm]
\small{E-mail: emilian.dudas@cpht.polytechnique.fr, palti@cpht.polytechnique.fr} \\[12mm]
\small{\bf Abstract} \\[3mm]
\end{center}
\begin{center}
\begin{minipage}[h]{16.0cm}
We study $SU(5)$ Grand Unified Theories within a local framework in F-theory with multiple extra $U(1)$ symmetries arising from a small monodromy group. The use of hypercharge flux for doublet-triplet splitting implies massless exotics in the spectrum that are protected from obtaining a mass by the $U(1)$ symmetries. We find that lifting the exotics by giving vacuum expectation values to some GUT singlets spontaneously breaks all the $U(1)$ symmetries which implies that proton decay operators are induced. If we impose an additional R-parity symmetry by hand we find all the exotics can be lifted while proton decay operators are still forbidden. These models can retain the gauge coupling unification accuracy of the MSSM at 1-loop. For models where the generations are distributed across multiple curves we also present a motivation for the quark-lepton mass splittings at the GUT scale based on a Froggatt-Nielsen approach to flavour.
\end{minipage}
\end{center}
\newpage
\setcounter{page}{1}
\pagestyle{plain}
\renewcommand{\thefootnote}{\arabic{footnote}}
\setcounter{footnote}{0}


\tableofcontents


\section{Introduction}
\label{sec:intro}

The idea that a Grand Unified Theory (GUT) underlies the Standard Model (SM) remains one of the most attractive ideas in theoretical particle physics. Perhaps the strongest phenomenological motivation for this is that once supersymmetry is introduced at the TEV scale the gauge couplings of the Minimal Supersymmetric Standard Model (MSSM) unify to a high accuracy at the GUT scale $2\times10^{16}$GEV. The unification of the coupling is certainly something we would like to retain in a GUT construction within string phenomenology. Recently string GUT constructions have been developed in the context of F-theory \cite{Donagi:2008ca,Beasley:2008dc,Beasley:2008kw,Donagi:2008kj} (see \cite{Heckman:2010bq} for a phenomenological review). One of the attractive features of these models is that they permit an elegant way of breaking the GUT group to the that of the standard model by turning on flux along the hypercharge direction in the unified gauge group \cite{Buican:2006sn,Donagi:2008kj,Beasley:2008kw}. The hypercharge flux however causes tension with gauge coupling unification. The most immediate problem arises from a direct non-universal correction to the gauge couplings at the GUT scale induced by the flux itself \cite{Blumenhagen:2008aw}. We do not address this issue and assume that this correction is small enough to keep within the coupling unification accuracy level or around 3 percent present in the MSSM. A more subtle problem arises from the fact that if the hypercharge flux is used to split the Higgs triplets from the doublets, it typically also induces massless exotics, by which we mean non-MSSM fields charged under the SM gauge groups \cite{Marsano:2009gv}.\footnote{This is always the case if one imposes a global $E_8$ structure. It may be possible to avoid exotics in models that are more general than those studied in this paper and we refer to section \ref{sec:exprot} for a discussion regarding this possibility.} In order to retain the MSSM gauge coupling unification these exotics must obtain mass. However they are usually protected against obtaining a mass by extra $U(1)$ symmetries typically present in such set-ups and can only gain a mass if some of these $U(1)$ symmetries are spontaneously broken. The problem is that these $U(1)$ symmetries are particularly useful for other purposes. For example to prevent a $\mu$-term or proton decay or to generate flavour hierarchies \cite{Bouchard:2009bu,Marsano:2009gv,Heckman:2009mn,Marsano:2009wr,Dudas:2009hu,King:2010mq}. It is the interplay between the exotics and the $U(1)$ symmetries that forms the primary motivation for this work and in particular the question: can we retain gauge coupling unification by lifting the exotic fields whilst still preventing a $\mu$-term and proton decay operators through the $U(1)$ symmetries?

In F-theory the GUT is realised on a 7-brane wrapping some 4-dimensional surface $S$ with other 7-branes intersecting $S$ along curves where matter representations are localised. Interactions between the matter representations are generally localised at points in $S$. One of the attractive features of such setups is that they can be studied within a local context. This means that a lot of information can be gained, for example regarding Yukawa couplings, simply by studying the local area around the point where the interaction is localised. In the case where all the matter interactions are localised on a single point such an approach can explore a wide range of model building aspects \cite{Heckman:2009mn,Dudas:2009hu,Heckman:2010xz}. Since the hypercharge flux is embedded within the GUT gauge group it is not localised at a point but rather over all of $S$. This means that aspects associated to GUT breaking, such as doublet-triplet splitting, are sensitive to the compact nature of $S$. It is still possible to study the properties of $S$ while decoupling the geometry of the full Calabi-Yau (CY) four-fold. This approach has been labeled semi-local and has been actively studied in \cite{Hayashi:2009ge,Donagi:2009ra,Marsano:2009gv,Marsano:2009wr,Hayashi:2010zp,Chen:2010tp}. We refer to \cite{Blumenhagen:2008zz,09024143,09060013,Blumenhagen:2009yv,Marsano:2009gv,Donagi:2009ra,09043932,Marsano:2009wr,Grimm:2009yu,Hayashi:2010zp,Grimm:2010ez,Marsano:2010ix} for fully global models.  In the semi-local approach the GUT theory is assumed to arise from a Higgsed $E_8$ gauge theory. The GUT theory arises from the breaking of
\be
E_8 \rightarrow SU(5)_{GUT}\times SU(5)_{\perp} \rightarrow SU(5)_{GUT}\times U(1)^4\;. \label{e8tosu5group}
\ee
The decoupling of the full CY geometry implies that some aspects must be chosen by hand and in particular the monodromy group experienced by the matter curves on $S$. This group identifies $U(1)$ factors in (\ref{e8tosu5group}) and also matter fields on curves related by the monodromy. The smaller the monodromy group the more $U(1)$s and matter curves remain. Semi-local constructions so far in the literature have only studied the case of a single $U(1)$ after the monodromy \cite{Marsano:2009gv,Marsano:2009wr}. We will study semi-local constructions with multiple $U(1)$s. There are two key motivations for this. The first is that having multiple $U(1)$s implies a better prospect for giving a mass to the exotics whilst retaining some protection against a $\mu$ term and proton decay. The second is that it is possible to realise the idea that each MSSM generation comes from a different matter curve which as shown in \cite{Dudas:2009hu,King:2010mq} can lead to attractive models of flavour.

In section \ref{sec:semloc}, following a brief review of some of the important geometric tools, we derive some general geometric properties of the curves on $S$ for the different possible monodromy groups that lead to multiple $U(1)$s. In particular we will calculate how the hypercharge flux restricts to the matter curves and also show that it is possible to have a semi-local realisation of such models without inducing exotic non-Kodaira singularities. With this geometric information we then proceed to perform some model building. In section \ref{sec:phenocon} we discuss the phenomenological constraints that are imposed on the models.  In section \ref{sec:singcur} we begin explicit model building in setups where all the MSSM generations are located on one matter curve. Following this in section \ref{sec:mulcur} we study models where the MSSM generations are located on different matter curves. In section \ref{sec:summary} we summarise our findings.

\section{Semi-local models with multiple $U(1)$s}
\label{sec:semloc}

In this section we develop semi-local constructions for small monodromy groups such that multiple $U(1)$ symmetries remain after imposing the monodromy identification. We begin by reviewing the basic ideas behind semi-local models in section \ref{sec:ellfib}. We then go on to construct new models with small monodromy groups. The final product for each construction is a determination of the restriction of the hypercharge flux to each of the matter curves.

It is important to stress that the models constructed in each subsection are not the most general setups that can be considered. They rely on a particular solution to the tracelessness constraints $b_1=0$ and they do not include all the possible monodromies. Further they rely on the assumption of an underlying $E_8$ singularity unfolded over the surface $S$ such that all curve and point enhacements on $S$ come from this single $E_8$. 

\subsection{The elliptic fibration}
\label{sec:ellfib}

The key properties of semi-local models are determined by how the elliptic fibration degenerates on the GUT surface $S$. We now review some of the relevant tools. For a small and incomplete selection of some of the original work and more recent pedagogical introductions see \cite{morrisonkatz, Bershadsky:1996nh, Katz:1996xe, 09024143, Hayashi:2009ge, Donagi:2009ra, Marsano:2009gv, Blumenhagen:2009yv}. Given the current extensive literature on the subject we shall be brief and only introduce the key concepts that will be used.

The Tate form of the elliptic fibration is given by
\be
x^3 - y^2 - xy \alpha_1 + x^2 \alpha_2 - y \alpha_3 + x \alpha_4 + \alpha_6 = 0 \;.
\ee
Here $(x,y)$ are affine coordinates on the torus fibre and the $\alpha_i$ are functions of the coordinates on the three-fold base.  The relevant data for the degeneration can be parameterised using the quantities
\bea
\beta_2 &=& \alpha_1^2 + 4\alpha_2 \;, \nn \\
\beta_4 &=& \alpha_1 \alpha_3 + 2\alpha_4 \;, \nn \\
\beta_6 &=& \alpha_3^2 + 4 \alpha_6 \;, \nn \\
\beta_8 &=& \beta_2 \alpha_6 - \alpha_1 \alpha_3 \alpha_4 + \alpha_2 \alpha_3^2 -\alpha_4^2 \;, \nn \\
\Delta &=& -\beta_2^2 \beta_8 - 8 \beta_4^3 - 27\beta_6^2 + 9 \beta_2 \beta_4 \beta_6 \;, \nn \\
f &=& -\frac{1}{48} \left(\beta_2^2 - 24 \beta_4 \right) \;, \nn \\
g &=& -\frac{1}{864} \left(-\beta_2^3 + 36\beta_2 \beta_4 - 216 \beta_6 \right) \;. \label{usequant}
\eea
Here $\Delta$ is the discriminant whose vanishing signals a singularity and $f$ and $g$ are defined as usual $\Delta = 4f^3 + 27g^2$. The type of singularity is determined by the order to which the discriminant vanishes as given by Kodaira's classification in table \ref{tab:kodsing}.
\begin{table}
\center
\begin{tabular}{|c|c|c|c|c|}
\hline
 ${\rm ord}(f)$ & ${\rm ord}(g)$ & ${\rm ord}(\Delta)$   & fiber type & singularity type \\
\hline
$\geq 0$ & $\geq 0$ & $0$ & smooth &none \\
\hline
$0$ & $0$ & $n$ &  $I_n$  & $A_{n-1}$ \\
\hline
$\geq 1$ & $1$ & $2$& $II$ & none\\
\hline
$1$ & $\geq 2$ & $3$ &  $III$ & $A_1$ \\
\hline
$\geq 2$ & $2$ & $4$ &   $IV$  & $A_2$\\
\hline
$2$ & $\geq 3$ & $n+6$ &  $I_n ^*$ & $D_{n+4}$ \\
\hline
$\geq 2$ & $3$ & $n+6$ &  $I_n ^*$ & $D_{n+4}$ \\
\hline
$\geq 3$ & $4$ & $8$ &  $IV^*$ & $E_6$ \\
\hline
$3$ & $\geq 5$ & $9$ &  $III^*$  & $E_7$\\
\hline
$\geq 4$ & $5$ & $10$ &  $II^*$  & $E_8$\\
\hline
\end{tabular}
\label{tab:kodsing}
\caption{Table showing Kodaira's classification of elliptic singularities.}
\end{table}

We are interested in an $SU(5)$ GUT that is localised on a divisor in the base three-fold. We choose the coordinates of the base so that the divisor is given by $z=0$. Then since we want an $SU(5)$ singularity on this divisor we can impose the order of vanishing of the $\alpha_i$ by writing them as
\be
\alpha_1 = b_5 \;,\;\; \alpha_2 = b_4 z \;,\;\; \alpha_3 = b_3 z^2 \;,\;\; \alpha_4 = b_2 z^3\;,\;\; \alpha_6 = b_0 z^5 \;, \label{su5sing}
\ee
where now the $b_i$s can depend on $z$ but do not vanish at $z=0$. For these it is easy to see that at $z=0$ we have $\ord(\beta_2)=0$ and so $\ord(f)=0$ and $\ord(g)=0$ while $\ord(\Delta)=5$ giving an $A_4$ singularity. The singularity is further enhanced when various combinations of the $b_i$ vanish. In particular using (\ref{su5sing}) we can write
\be
\Delta = - z^5 \left[P^4_{10} P_5 + z P^2_{10} \left(8b_4P_5 + b_5 R\right) + {\cal O}\left(z^2\right)  \right] \;,
\ee
where we define\footnote{Also $R=-b_3^3-b_2^2b_5 + 4b_0b_4b_5$.}
\bea
P_{10} &=& b_5 \;, \\
P_5 &=& b_3^2 b_4 - b_2 b_3 b_5 + b_0 b_5^2 \;. \label{p5p10}
\eea
This means that, on the GUT divisor $z=0$, if $P_5=0$ the singularity enhances to at least $SU(6)$ and if $P_{10}=0$ the singularity enhances to at least $SO(10)$. These loci correspond to the curves on the GUT divisor $S$ on which matter in the $\f\oplus\fb$ and $\te\oplus\teb$ representations is localised as can be determined by decomposing the adjoint of the enhanced gauge group under $SU(5)$ representations. We denote these as $\f$-matter curves and $\te$-matter curves respectively. For other vanishing combinations it is possible to enhance further corresponding to intersections of the matter curves. In particular the point of $E_8$ discussed in the introduction corresponds to $b_{i\neq0}=0$.

In a global model the determination of the $b_i$'s as functions of the base coordinates gives the structure of the GUT theory. However in a semi-local approach it is possible to bypass some of the complications by looking close to the GUT divisor. In this approach we can consider the full CY four-fold as given by an Asymptotically Locally Euclidean (ALE) space with ADE singularities fibered over the GUT divisor. This can be modeled by considering an $E_8$ singularity in the fibre which is resolved by blowing up the collapsed two-cycles as we move around the GUT divisor. Since we can write $E_8=SU(5)_{GUT}\times SU(5)_{\perp}$ the two-cycles correspond to generators in $SU(5)_{\perp}$ and so we write the curves on which they collapse using the $t_i$ with $i=1,..,5$ and $\sum t_i =0$. Explicitly using the decomposition of the adjoint of $E_8$
\be
\bf{248} \rightarrow \left(\bf{24},\bf{1}\right)\op\left(\bf{1},\bf{24}\right)\op\left(\te,\f\right)\op\left(\fb,\te\right)\op\left(\teb,\fb\right)\op\left(\f,\teb\right) \;,
\ee
we see that we have 5 $\te$-matter curves, 10 $\f$-matter curves and 24 singlets which can be parameterised by the vanishing combinations
\bea
\Sigma_{\te\op\teb} &:& t_i = 0 \;, \label{cur1} \\
\Sigma_{\f\op\fb} &:& -t_i - t_j = 0, \;\; i\neq j\;, \label{cur2} \\
\Sigma_{\bf{1}} &:& \pm \left(t_i - t_j\right) = 0 , \;\; i\neq j\;. \label{cur3}
\eea
The $b_i$ are given in terms of elementary symmetric polynomials of degree $i$ in the $t_i$.
An important point is that since this is a non-linear relation there can be branch cuts that connect the various $t_i$ and these are the monodromies discussed in the introduction. If some $t_i$ lie in the same orbit of the monodromy group they can, for the purposes of this analysis, be identified. This reduces the number of matter curves and also the number of $U(1)$ gauge symmetries.

A useful way to encode all this information is by using the spectral cover construction. More precisely we need to introduce the spectral cover for the fundamental representations of $SU(5)_{\perp}$. The spectral cover is a hypersurface inside the projective 3-fold
\be
X = {\mathbb P}({\cal O}_{S_{GUT}}\oplus K_{S_{GUT}}) \;,
\ee
given by the constraint
\be
C_{10} = b_0 U^5 + b_2 V^2 U^3 + b_3 V^3 U^2 + b_4 V^4 U + b_5 V^5 = 0 \;. \label{spect10}
\ee
Here ${\cal O}_{S_{GUT}}$ and $K_{S_{GUT}}$ are the trivial and canonical bundle on $S_{GUT}$ respectively and $\{U,V\}$ are homogeneous complex coordinates on the ${\mathbb P}^1$ fibre in $X$. The idea is that locally we can set some affine parameter $s=U/V$ in which (\ref{spect10}) is a polynomial whose 5 roots are exactly the $t_i$. Indeed $s$ can be equated with the value of the Higgs field that breaks the $E_8$ gauge theory and overall (\ref{spect10}) forms a 5-fold cover of $S_{GUT}$. The monodromy of the Higgs or the $t_i$ is encoded in the global properties of (\ref{spect10}) and more specifically in how the polynomial decomposes into products. We can think of all the $\te$-matter curves as lifting to a single curve on the spectral cover which then decomposes into parts according to the decomposition of the spectral cover. Indeed this curve is determined by the equation $U=0$ which gives
\be
P_{10} = b_5 = t_1 t_2 t_3 t_4 t_5 = 0 \;,
\ee
which reproduces the equations for the 5 $\te$-matter curves.

Having introduced the necessary tools we go on to study the structure of the matter curves for various monodromy groups. There are only 3 types of monodromy actions that preserve at least 2 independent $U(1)$s. We denote them by which $t_i$s are identified or equivalently how the spectral cover factorises. The 3 cases are for factorisations of the type $2+2+1$, $3+1+1$ and $2+1+1+1$, where for example the first case denotes identifying $t_1$ with $t_2$ and $t_3$ with $t_4$.\footnote{We do not consider cases where the monodromy group is a subgroup of the factorisation which can only occur for factors of degree 4 or 5 \cite{Marsano:2009gv}. Also we do not consider cases with a single $U(1)$ corresponding to factorisation $3+2$ and $4+1$ which have been studied in  \cite{Marsano:2009gv,Marsano:2009wr}.}

\subsection{Matter curves for a ${2+1+1+1}$ splitting}
\label{sec:matcur2111}

In this section we consider the case where the spectral cover decomposes into 4 pieces
\be
C_{10} = \left(a_1 V^2 + a_2 V U + a_3 U^2 \right) \left(a_4 V + a_7 U\right) \left(a_5 V + a_8 U\right) \left(a_6 V + a_9 U\right) = 0 \;. \label{spect10z2}
\ee
Here the $a_I$ are some as yet undetermined coefficients that are functions on $S$. This decomposition corresponds to a ${\mathbb Z}_2$ monodromy group that by choice of parameterisation we shall take to act as $t_1 \leftrightarrow t_2$. So that the $\te$-curves $t_1$ and $t_2$ both lift to a curve on a single factor of the spectral cover given by the first brackets in (\ref{spect10z2}). We can write the $b_i$ in terms of the $a_I$ as
\bea
b_0 &=& a_{3789} \;, \nn \\
b_1 &=& a_{2789} + a_{3678} + a_{3579} + a_{3489} \;, \nn \\
b_2 &=& a_{1789} + a_{2678} + a_{2579} + a_{2489} + a_{3567} + a_{3468} + a_{3459} \;, \nn \\
b_3 &=& a_{3456} + a_{1678} + a_{1579} + a_{1489} + a_{2567} + a_{2468} + a_{2459} \;, \nn \\
b_4 &=& a_{2456} + a_{1567} + a_{1468} + a_{1459} \;, \nn \\
b_5 &=& a_{1456} \;. \label{aIsol2111}
\eea
Here we use the notation $a_{IJKL}=a_I a_J a_K a_L$.

We are interested in determining the curves $a_I=0$ on $S_{GUT}$. This can be done as follows. The $b_i$ are zero sections of the bundle $\eta - i c_1$ \cite{Donagi:2009ra,Marsano:2009gv}. Here $c_1$ is the first Chern class of the tangent bundle of $S_{GUT}$ and $\eta = 6 c_1 - t$ with $-t$ being the first Chern class of the normal bundle to $S_{GUT}$. Using (\ref{aIsol2111}) this then implies that the $a_I$ are sections of bundles as shown in table \ref{tab:aIsect2111}.\footnote{Note that the $\eta$ factor is not completely determined by the $b_i$ and can be chosen such that all the sections are sufficiently positive. Generally, for all the monodromy groups, taking $\eta$ and the $\chi$ sufficiently positive means the $a_i$ are holomorphic sections.}
\begin{table}
\centering
\begin{tabular}{|c|c|}
\hline
Section & $c_1$(Bundle)\\
\hline
$a_1$ & $\eta - 2c_1 - \tilde{\chi}$ \\
\hline
$a_2$ & $\eta - c_1 - \tilde{\chi}$ \\
\hline
$a_3$ & $\eta - \tilde{\chi}$\\
\hline
$a_4$ & $-c_1 + \chi_7$ \\
\hline
$a_5$ & $-c_1 + \chi_8$\\
\hline
$a_6$ & $-c_1 + \chi_9$\\
\hline
$a_7$ & $\chi_7$ \\
\hline
$a_8$ & $\chi_8$\\
\hline
$a_9$ & $\chi_9$\\
\hline
\end{tabular}
\caption{Table showing the first Chern classes of the line bundles that the $a_I$ are sections of for the factorisation $2+1+1+1$. The forms $\chi_{\{7,8,9\}}$ are unspecified and we define $\tilde{\chi}=\chi_7 + \chi_8 + \chi_9$.}
\label{tab:aIsect2111}
\end{table}
Here $\chi_{\{7,8,9\}}$ are unspecified and we define $\tilde{\chi}=\chi_7 + \chi_8 + \chi_9$.

We can gain more information by noting that since
\be
b_1 = t_1 + t_2 + t_3 + t_4 + t_5 = 0 \;,
\ee
we have the constraint on the $a_i$ that
\be
a_2a_7a_8a_9 + a_3a_6a_7a_8 + a_3a_5a_7a_9 + a_3a_4a_8a_9 = 0\;. \label{traceconst}
\ee
There are a number of ways to solve this constraint but, as also noted in \cite{Marsano:2009wr}, most lead to non-Kodaira type singularities on the manifold. By this we mean that over some curves/points on $S$ the $b_i$ vanish to such an order that a singularity is induced that does not fall into the classification of table \ref{tab:kodsing}. However we can take the following ansatz
\bea
a_2 &=& - c \left( a_6 a_7 a_8 + a_5 a_7 a_9 + a_4 a_8 a_9\right) \;, \nn \\
a_3 &=& c\; a_7 a_8 a_9\;, \label{b1sol2111}
\eea
where $c$ is some unspecified holomorphic section in the homology class
\be
[c] = \eta - 2 \tilde{\chi}\;.
\ee
Of course this assumes that in a global model such a constraint can be met which is a non-trivial assumption. With this choice we can write
\bea
b_0 &=& c \left(a_7a_8a_9\right)^2 \;, \nn \\
b_1 &=& 0 \;, \nn \\
b_2 &=& a_1a_7a_8a_9 - c\left[ \left(a_6a_7a_8\right)^2 + a_6a_7a_8a_9\left(a_5a_7+a_4a_8\right) + a^2_9\left(a_5^2a_7^2 + a_4a_5a_7a_8 + a_4^2a_8^2  \right) \right] \;, \nn \\
b_3 &=& a_1\left(a_6 a_7 a_8 + a_5 a_7 a_9 + a_4 a_8 a_9 \right) - c\left(a_5 a_7 + a_4 a_8 \right) \left(a_6 a_7 + a_4 a_9 \right) \left(a_6 a_8 + a_5 a_9 \right) \;, \nn \\
b_4 &=& a_1\left(a_5 a_6 a_7 + a_4 a_6 a_8 + a_4 a_5 a_9 \right) - ca_4a_5a_6\left(a_6 a_7 a_8 + a_5 a_7 a_9 + a_4 a_8 a_9\right) \;, \nn \\
b_5 &=& a_1a_4a_5a_6 \;,  \label{bisol2111}
\eea
We can check that this does not lead to any exotic non-Kodaira singularities. For example consider setting $a_1=0$ then using the results of section \ref{sec:ellfib} we find a Kodaira singularity of $SO(10)$ signaling a $\te$-matter curve.\footnote{The procedure is to use (\ref{bisol2111}) to read off the vanishing order of the $b_i$, which then give the vanishing order of the $\alpha_i$ through (\ref{su5sing}) which then can be used to determine the vanishing order of $f$, $g$ and $\Delta$ through (\ref{usequant}) which then give the singularity type as in table \ref{tab:kodsing}.} The same follows for $a_4$, $a_5$ and $a_6$. We can consider $a_7=0$ which gives an $A_4$ singularity and the same for $a_8$ and $a_9$ and $c$. It is possible to generate bad singularities say if $a_1=c=0$ but then this just implies that these curves should not intersect.

After imposing (\ref{b1sol2111}) the $\f$-matter curve polynomial (\ref{p5p10}) decomposes as
\bea
P_5 &=& \left(a_5 a_7 + a_4 a_8 \right) \left(a_6 a_7 + a_4 a_9 \right) \left(a_6 a_8 + a_5 a_9 \right) \nn \\
 & & \left( a_6 a_7 a_8 + a_5 a_7 a_9 + a_4 a_8 a_9\right) \nn \\
 & & \left( a_1 - c a_5 a_6 a_7 - c a_4 a_6 a_8 \right) \nn \\
 & & \left( a_1 - c a_5 a_6 a_7 - c a_4 a_5 a_9 \right) \nn \\
 & & \left( a_1 - c a_4 a_6 a_8 - c a_4 a_5 a_9 \right) \;. \label{Z2P5curves}
\eea
These are the 7 $\f$-matter curves that are left after the ${\mathbb Z}_2$ monodromy. We describe these curves in table \ref{tab:curves2111}.
\begin{table}[ht]
\center
{\small
\begin{tabular}{|cccccc|}
\hline
Matter & Charge & Equation & Homology & $N_{\mathrm{Y}}$ & $M_{U(1)}$ \\
\hline
$\f_{H_u}$  & $-2t_1$    & $a_6 a_7 a_8 + a_5 a_7 a_9 + a_4 a_8 a_9$ & $-c_1 + \tilde{\chi}$         & $\tilde{N}$   & $M_{\f_{H_u}}$ \\
$\f_{1}$    & $-t_1-t_3$ & $a_1 - c a_4 a_6 a_8 - c a_4 a_5 a_9$     & $\eta - 2c_1 - \tilde{\chi}$  & $-\tilde{N}$  & $M_{\f_1}$ \\
$\f_{2}$    & $-t_1-t_4$ & $a_1 - c a_5 a_6 a_7 - c a_4 a_5 a_9$     & $\eta - 2c_1 - \tilde{\chi}$  & $-\tilde{N}$  & $M_{\f_2}$ \\
$\f_{3}$    & $-t_1-t_5$ & $a_1 - c a_5 a_6 a_7 - c a_4 a_6 a_8$     & $\eta - 2c_1 - \tilde{\chi}$  & $-\tilde{N}$  & $M_{\f_3}$ \\
$\f_{4}$    & $-t_3-t_4$ & $a_5 a_7 + a_4 a_8$                       & $-c_1 + \chi_7 + \chi_8$      & $N_7+N_8$                        & $M_{\f_4}$ \\
$\f_{5}$    & $-t_3-t_5$ & $a_6 a_7 + a_4 a_9$                       & $-c_1 + \chi_7 + \chi_9$      & $N_7+N_9$                        & $M_{\f_5}$ \\
$\f_{6}$    & $-t_4-t_5$ & $a_6 a_8 + a_5 a_9$                       & $-c_1 + \chi_8 + \chi_9$      & $N_8+N_9$                        & $M_{\f_6}$ \\
$\te_{M}$   & $t_1$      & $a_1$                                     & $\eta - 2c_1 -\tilde{\chi}$   & $- \tilde{N}$ & $-\left(M_{\f_1}+M_{\f_2}+M_{\f_3}\right)$ \\
$\te_{2}$   & $t_3$      & $a_4$                                     & $-c_1 + \chi_7$               & $N_7$                            & $M_{\te_2}$ \\
$\te_{3}$   & $t_4$      & $a_5$                                     & $-c_1 + \chi_8$               & $N_8$                            & $M_{\te_3}$ \\
$\te_{4}$   & $t_5$      & $a_6$                                     & $-c_1 + \chi_9$               & $N_9$                            & $M_{\te_4}$ \\
\hline
\end{tabular}
}
\caption{Table showing curves and flux restrictions for $2+1+1+1$ splitting. We have defined $\tilde{N}=N_7+N_8+N_9$.}
\label{tab:curves2111}
\end{table}
Note that 3 of the $\f$-matter curves share the same homology class $\left[\f_{1} \right]=\left[\f_{2} \right]=\left[\f_{3}\right]$. This implies that any flux restricts to them in the same way. It also implies that their intersections are determined by the number of self-intersections.

Having determined the homology classes of the matter curves we can determine the induced chiral spectrum in terms of the restriction of the fluxes to the curves. There are two types of fluxes that contribute to the spectrum. The first is flux turned on in the 4 $U(1)$s, or rather the number of $U(1)$s left after the monodromy identification which in the case of ${\mathbb Z}_2$ is 3, of the $SU(5)_{\perp}$. This flux respects the $SU(5)$ GUT structure and so only affects the chirality of complete GUT multiplets. We refer to this type of flux henceforth as $U(1)$-flux and denote it by $F_{U(1)}$. The second type of flux is turned on along the hypercharge direction in $SU(5)_{GUT}$. This flux determines the splitting of the GUT matter multiplets. We refer to this type of flux henceforth as hypercharge flux and denote it by $F_Y$. Given a restriction to a curve of the $U(1)$ flux given by an integer $M$ and the hypercharge flux given by an integer $N$ we have the spectrum \cite{Donagi:2008ca,Beasley:2008dc,Beasley:2008kw,Donagi:2008kj,Marsano:2009wr,Blumenhagen:2009yv}\footnote{More precisely the flux is specified by fractional line-bundles $L_{Y}$, $V_{10}$ and $V_{5}$ (sometimes denoted as $V$ and $\wedge^2V$ respectively) such that $M_{10}=\mathrm{deg}\left(L_{Y}^{1/6}\otimes V_{10}\right)$, $M_{5}=\mathrm{deg}\left(L_{Y}^{-1/3}\otimes V_{5}\right)$ and $N=\mathrm{deg}\left(L_{Y}^{5/6}\right)$.}
\bea
n_{(3,1)_{-1/3}} - n_{(\bar{3},1)_{+1/3} } &=& M_{5} \;, \nn \\
n_{(1,2)_{+1/2}} - n_{(1,2)_{-1/2} } &=& M_{5} + N \;,
\eea
for the 5-matter curves and
\bea
n_{(3,2)_{+1/6}} - n_{(\bar{3},2)_{-1/6} } &=& M_{10} \;, \nn \\
n_{(\bar{3},1)_{-2/3}} - n_{(3,1)_{+2/3} } &=& M_{10} - N \;, \nn \\
n_{(1,1)_{+1}} - n_{(1,1)_{-1} } &=& M_{10} + N\;,
\eea
for the 10-matter curves.

Since the $U(1)$ fluxes are turned on along the world-volume of branes that are not restricted to $S_{GUT}$ but rather probe the full geometry of the CY four-fold their determination requires knowledge of the full compact geometry. Therefore for our purposes we shall take their restriction to the matter curves as free parameters. This ignores any subtleties to do with quantisation conditions and other issues that may come up in a global context. We note also that some information on the $U(1)$ flux can be gained in a semi-local context using the so called universal flux \cite{Donagi:2009ra,Marsano:2009wr} but a complete study requires also non-universal fluxes which are constructed in a global model. There are some mild restrictions that can be imposed locally since whatever form the flux takes it restricts to elements in the same homology class identically. Combining this with the tracelessness condition, i.e. $\sum_i F_{U(1)_i} = 0$, gives the universal constraint
\be
\sum M_{\te} = - \sum M_{\f} \;,
\ee
which is just anomaly cancellation. Further we also find the relation between $M_{\te_1}$ and $M_{\f_1}+M_{\f_2}+M_{\f_3}$ shown in table \ref{tab:curves2111}.

The hypercharge flux on the other hand is restricted purely to $S_{GUT}$ and so is more constrained within a semi-local model. In particular we require that it restricts trivially to any matter curves on $S_{GUT}$ that are lifted to non-trivial homology classes of the full CY four-fold. This is of course the requirement that it does not receive a Green-Schwarz mass \cite{Buican:2006sn,Beasley:2008kw,Donagi:2008kj}. In terms of the introduced homology classes such restriction translates to \cite{Marsano:2009gv}
\be
F_Y \cdot c_1 = 0 \;, \;\; F_Y \cdot \eta = 0 \;.
\ee
Therefore for the ${\mathbb Z}_2$ monodromy model we have the restrictions as in table \ref{tab:curves2111}. In particular, as pointed out in \cite{Marsano:2009gv}, if the hypercharge restricts non-trivially to any $\f$-matter curves it must also restrict non-trivially to a $\te$-matter curve. This implies that using the hypercharge flux for doublet-triplet splitting, as suggested in \cite{Beasley:2008kw}, implies that some non-GUT exotics appear. Note also that the sum over the hypercharge flux for each type of matter curve vanishes which means that the hypercharge flux induces no net chirality overall.

The flux restrictions in table \ref{tab:curves2111} are the required data to begin model building. This essentially amounts to picking $M$s and $N$s freely and studying the resulting phenomenology. This is the subject of sections \ref{sec:singcur} and \ref{sec:mulcur} but before proceeding we perform a similar analysis for the other possible factorisations.

\subsection{Matter curves for a ${2+2+1}$ splitting}
\label{sec:matcur221}

Since in the previous section we studied the factorisation in detail in the next two sections we briefly state the results without repeating the discussions of the calculations. In the $2+2+1$ case we have the spectral cover splitting as
\be
C_{10} = \left(a_1 v^2 + a_2vu +a_3u^2\right)\left(a_4 v^2 + a_5vu +a_6u^2\right)\left(a_7 v + a_8u\right) \;.
\ee
The $b_i$ are given by
\bea
b_0 &=& a_{368} \;, \nn \\
b_1 &=& a_{367} + a_{358} + a_{268}\;, \nn \\
b_2 &=& a_{357} + a_{267} + a_{348} + a_{258} + a_{168}\;, \nn \\
b_3 &=& a_{347} + a_{257} + a_{167} + a_{248} + a_{158}\;, \nn \\
b_4 &=& a_{247} + a_{157} + a_{148} \;, \nn \\
b_5 &=& a_{147} \;. \label{aIsol221}
\eea
This then implies that the $a_I$ transform as shown in table \ref{tab:aIsect221}.
\begin{table}
\centering
\begin{tabular}{|c|c|}
\hline
Section & $c_1$(Bundle)\\
\hline
$a_1$ & $-2c_1 + \chi_1$ \\
\hline
$a_2$ & $- c_1 + \chi_1$ \\
\hline
$a_3$ & $\chi_1$\\
\hline
$a_4$ & $-2c_1 + \chi_2$ \\
\hline
$a_5$ & $-c_1 + \chi_2$\\
\hline
$a_6$ & $\chi_2$\\
\hline
$a_7$ & $\eta-c_1-\chi_1-\chi_2$ \\
\hline
$a_8$ & $\eta-\chi_1-\chi_2$\\
\hline
\end{tabular}
\caption{Table showing the first Chern classes of the line bundles that the $a_I$ are sections of for the factorisation $2+2+1$.}
\label{tab:aIsect221}
\end{table}
We solve the $b_1=0$ constraint by the following ansatz\footnote{There is an equivalent possibility taking $a_2\leftrightarrow a_5$ and $a_3\leftrightarrow a_6$ which just amounts to relabeling.}
\bea
a_2 &=& - c \left( a_6 a_7 + a_5 a_8\right) \;, \nn \\
a_3 &=& c\; a_6 a_8 \;,
\eea
where $c$ is some unspecified holomorphic section in the homology class
\be
[c] = -\eta + 2 \chi_1\;. \label{secc}
\ee
With this choice we find
\bea
b_0 &=& a_6^2a_8^2c \;, \nn \\
b_1 &=& 0 \;, \nn \\
b_2 &=& a_1a_6a_8 + c\left(-a_6^2a_7^2 - a_5a_6a_7a_8 - a_5^2a_8^2 + a_4a_6a_8^2 \right) \;, \nn \\
b_3 &=& a_1a_6a_7 + a_1a_5a_8 - c\left( a_5a_6a_7^2 + a_5^2a_7a_8 + a_4a_5a_8^2\right)\;, \nn \\
b_4 &=& a_1a_5a_7 + a_1a_4a_8 - c\left(a_4a_6a_7^2 + a_4a_5a_7a_8\right) \;, \nn \\
b_5 &=& a_1a_4a_7 \;.
\eea
This does not lead to any non-Kodaira singularities. The $P_5$ polynomial decomposes into the product of the polynomials given in table \ref{tab:curves221} where also the relevant data is summarised.
\begin{table}[ht]
\center
{\small
\begin{tabular}{|cccccc|}
\hline
Matter & Charge & Equation & Homology & $N_{\mathrm{Y}}$ & $M_{U(1)}$ \\
\hline
$\f_{H_u}$  & $-2t_1$    & $a_6a_7+a_5a_8$                           & $\eta-c_1 - \chi_1$         & $-N_1$   & $M_{\f_{H_u}}$ \\
$\f_{1}$    & $-t_1-t_3$ & $\begin{array}{c} a_1^2 - a_1a_5a_7 c - 2a_1a_4a_8c \\ + a_4a_6a_7^2c^2 + a_4a_5a_7a_8c^2 + a_4^2a_8^2c^2\end{array}$     & $-4c_1 + 2\chi_1$  & $2N_1$  & $M_{\f_1}$ \\
$\f_{2}$    & $-t_1-t_5$ & $a_1 - a_5a_7c$     & $- 2c_1 + \chi_1$  & $N_1$  & $M_{\f_2}$ \\
$\f_{3}$    & $-t_3-t_5$ & $a_6a_7^2 + a_5a_7a_8 + a_4a_8^2$     & $2\eta - 2c_1 - 2\chi_1 -\chi_2$  & $-2N_1-N_2$  & $M_{\f_3}$ \\
$\f_{4}$    & $-2t_3$ & $a_5$                       & $-c_1 + \chi_2$      & $N_2$                        & $M_{\f_4}$ \\
$\te_{M}$   & $t_1$      & $a_1$                                     & $- 2c_1 +\chi_1$   & $N_1$ & $-\left( M_{\f_1}+M_{\f_2} \right)$ \\
$\te_{2}$   & $t_3$      & $a_4$                                     & $-2c_1 + \chi_2$               & $N_2$                            & $M_{\te_2}$ \\
$\te_{3}$   & $t_5$      & $a_7$                                     & $\eta-c_1 - \chi_1 - \chi_2$               & $-N_1-N_2$                            & $M_{\te_3}$ \\
\hline
\end{tabular}
}
\caption{Table showing curves for $2+2+1$ splitting.}
\label{tab:curves221}
\end{table}

\subsection{Matter curves for a ${3+1+1}$ splitting}
\label{sec:matcur311}

In this case we have the spectral cover splitting as
\be
C_{10} = \left(a_1 v^3 + a_2v^2u +a_3vu^2 + a_4u^3\right)\left(a_5 v + a_6u\right)\left(a_7 v + a_8u\right) \;.
\ee
The $b_i$ are given by
\bea
b_0 &=& a_{468} \;, \nn \\
b_1 &=& a_{467} + a_{458} + a_{368}\;, \nn \\
b_2 &=& a_{457} + a_{367} + a_{358} + a_{268}\;, \nn \\
b_3 &=& a_{357} + a_{267} + a_{258} + a_{168}\;, \nn \\
b_4 &=& a_{257} + a_{167} + a_{158} \;, \nn \\
b_5 &=& a_{157} \;. \label{aIsol311}
\eea
This then implies that the $a_I$ transform as shown in table \ref{tab:aIsect311}.
\begin{table}
\centering
\begin{tabular}{|c|c|}
\hline
Section & $c_1$(Bundle)\\
\hline
$a_1$ & $\eta- 3c_1 - \left(\chi_1+\chi_2\right)$ \\
\hline
$a_2$ & $\eta- 2c_1 - \left(\chi_1+\chi_2\right)$ \\
\hline
$a_3$ & $\eta- c_1 - \left(\chi_1+\chi_2\right)$\\
\hline
$a_4$ & $\eta - \left(\chi_1+\chi_2\right)$ \\
\hline
$a_5$ & $-c_1 + \chi_1$\\
\hline
$a_6$ & $\chi_1$\\
\hline
$a_7$ & $-c_1+\chi_2$ \\
\hline
$a_8$ & $\chi_2$\\
\hline
\end{tabular}
\caption{Table showing the first Chern classes of the line bundles that the $a_I$ are sections of for the factorisation $3+1+1$.}
\label{tab:aIsect311}
\end{table}
We solve the $b_1=0$ constraint by the following ansatz
\bea
a_3 &=& - c \left( a_6 a_7 + a_5 a_8\right) \;, \nn \\
a_4 &=& c\; a_6 a_8 \;,
\eea
where $c$ is some unspecified holomorphic section in the homology class
\be
[c] = \eta - 2 \left(\chi_1 + \chi_2\right)\;.
\ee
With this choice we find
\bea
b_0 &=& a_6^2a_8^2c \;, \nn \\
b_1 &=& 0 \;, \nn \\
b_2 &=& a_2a_6a_8 - c\left(a_6^2a_7^2 + a_5a_6a_7a_8 + a_5^2a_8^2 \right) \;, \nn \\
b_3 &=& a_1a_6a_8 + a_2a_5a_8 +a_2a_6a_7 - c\left( a_5a_6a_7^2 + a_5^2a_7a_8\right)\;, \nn \\
b_4 &=& a_2a_5a_7 + a_1a_6a_7 + a_1a_5a_8 \;, \nn \\
b_5 &=& a_1a_5a_7 \;, \nn \\
\eea
This does not lead to any non-Kodaira singularities. The $P_5$ polynomial decomposes into the product of the polynomials given in table \ref{tab:curves311} where also the relevant data is summarised.
\begin{table}[ht]
\center
{\small
\begin{tabular}{|cccccc|}
\hline
Matter & Charge & Equation & Homology & $N_{\mathrm{Y}}$ & $M_{U(1)}$ \\
\hline
$\f_{H_u}$  & $-2t_1$    & $a_2a_6a_7+a_2a_5a_8+a_1a_6a_8$ & $\eta-3c_1$         & $0$   & $M_{\f_{H_u}}$ \\
$\f_{1}$    & $-t_1-t_4$ & $a_2a_5+a_1a_6-a_5^2a_7c$     & $\eta - 3c_1 - \chi_2$  & $-N_2$  & $M_{\f_1}$ \\
$\f_{2}$    & $-t_1-t_5$ & $a_2a_7+a_1a_8-a_5a_7^2c$     & $\eta - 3c_1 - \chi_1$  & $-N_1$  & $M_{\f_2}$ \\
$\f_{3}$    & $-t_4-t_5$ & $a_6a_7+a_5a_8$     & $-c_1+\chi_1+\chi_2$  & $N_1+N_2$  & $M_{\f_3}$ \\
$\te_{M}$   & $t_1$      & $a_1$                                     & $\eta - 3c_1 -\chi_1-\chi_2$   & $-N_1-N_2$ & $M_{\te_M}$ \\
$\te_{2}$   & $t_4$      & $a_5$                                     & $-c_1 + \chi_1$               & $N_1$                            & $M_{\te_2}$ \\
$\te_{3}$   & $t_5$      & $a_7$                                     & $-c_1 + \chi_2$               & $N_2$                            & $M_{\te_3}$ \\
\hline
\end{tabular}
}
\caption{Table showing curves for $3+1+1$ splitting.}
\label{tab:curves311}
\end{table}

\section{Phenomenological constraints}
\label{sec:phenocon}

In this section we discuss in general terms the key phenomenological aspects of the models. In particular the tension between lifting the exotics and preventing proton decay.

\subsection{The exotics mass and proton decay}
\label{sec:exprot}

It was shown in \cite{Marsano:2009gv} that the use of hypercharge flux for doublet-triplet splitting typically induces exotics that are not in complete GUT multiplets. The more precise statement is that if one imposes a global $E_8$ structure over all of $S$, such that any singularity enhancements are associated to this global $E_8$, then exotics are always induced by hypercharge flux. This includes all models with heterotic duals. In general it may be possible to avoid this in models where there are enhancements not associated to a single global $E_8$ but such models are not of the semi-local type studied in this paper. Note further that the existence of a global $E_8$ is not implied by the existence of a point of $E_8$ enhancement as studied in \cite{Heckman:2009mn,Dudas:2009hu}.
An important fact is that because the hypercharge flux is required to be trivial in homology in order to not gain a Green-Schwarz mass \cite{Buican:2006sn,Beasley:2008kw,Donagi:2008kj} it does not induce any net chirality with respect to the GUT gauge group. This means that the exotics always come in vector pair representations. However they are still forbidden from obtaining a mass by the extra $U(1)$ symmetries inherited from the $E_8$ structure. The exotics do have renormalisable couplings to GUT singlets which can be used to give them a mass through a vacuum expectation value. So the mass terms for the exotics take the form
\be
W \supset X {\bf R} {\bf \bar{R}} \;,
\ee
where $X$ denotes a GUT singlet, ${\bf R}$ denotes a SM component of a GUT representation with ${\bf \bar{R}}$ its conjugate.\footnote{It may be that we have multiple matter generations on one curve and so we should consider whether all of them can be lifted by such an interaction. For example consider the curve holding the ${\bf R}$ representation to hold $I$ generations and that of ${\bf \bar{R}}$ to hold $J$ with $J<I$. Then if the mass matrix takes the most general form we expect that by chirality $J$ vector pairs of fields are lifted by such an interaction leaving $I-J$ massless modes. This means that, for example, if on a curve holding 3 SM generations also there is an exotic then we count this as 1 exotic field and given a mass interaction with a curve holding a single generation of the chiral conjugate representation we take the exotic to be lifted. More precisely what would happen is that the remaining 3 massless modes will be linear combinations of all 4 generations on the SM curve. There is a subtlety to do with whether the mass matrix does take the most general form. The reason is that there may be local $U(1)$ symmetries that prevent the interactions as occurs for Yukawa couplings where, in the absence of non-commutative deformations, the matrix is exactly rank 1 \cite{Heckman:2008qa,Font:2009gq,Cecotti:2009zf,Conlon:2009qq}. However in the case where the singlet is participating, since its wavefunction is not localised on $S$, such a cancellation seems less likely \cite{Bouchard:2009bu}. We leave a more through study of this effect for future work and for now assume that the mass matrix is general enough for exotics to be lifted unless protected by chirality.}
The problem with this is that the singlets vev spontaneously break some of the $U(1)$ symmetries. Indeed generically they should break them quite strongly since in order to retain gauge coupling unification the exotics should have a large mass. Non-generically it is possible to only break them slightly if the exotics act as complete GUT multiplets in the beta functions since then gauge coupling unification is unaltered at 1-loop. Breaking the $U(1)$ symmetries implies that we should re-examine whether they can be used to solve some of the problems of GUT theories which we now turn to.

\subsubsection*{Constraints on proton decay operators}

A famous problem with minimal $SU(5)$ GUT theories is that they predict heavy Higgs triplets and such triplets can mediate proton decay which is very constrained. Since we are working with non-minimal GUT theories it is worth recalling first what the constraints are directly on proton decay operators.
There are two types of proton decay operators of dimensions 4 and 5 (we do not consider dimension 6 proton decay)
\be
\lambda \fb_{M}\fb_{M}\te_{M} \;,\;\; W \te_M \te_M \te_M \fb_M \;.
\ee
Here the subscript $M$ refers to the curve on which the matter representations are localised which means any generation. The dimension 4 operator is constrained for any generation indices to be $\lambda_{ijk}<10^{-5}$ \cite{Smirnov:1996bg}. The dimension 5 operator is less clear. Studies of constraints on such operators were done in \cite{salatil,Ellis:1983qm,Ibanez:1991pr,Barbier:2004ez}. There the limit on the particular generation structure was given as $W_{112l}<\frac{10^{-10}}{M_{GUT}}$. Here subscript $1$ refers to the lightest generation and $l$ refers to either of the two light leptons. With updated proton decay lifetime constraints, which have prolonged the lifetime by around $10^3$ we expect this limit to be increased by $10^1-10^2$ so that as a crude estimate
\be
W_{112l}<\frac{10^{-11}}{M_{GUT}}. \label{dim5prot}
\ee
We have used the suppression scale of $M_{GUT}$ which follows from the results of \cite{Conlon:2007zza} which show that the correct suppression scale should be the `winding' scale $M_s {\cal V}^{1/6}$, where ${\cal V}$ is the $B_6$ or CY 3-fold volume, and $M_s$ is the string scale. This is combined with the results of \cite{Conlon:2009qa,Conlon:2009xf,Conlon:2009kt} which show that this is also the unification scale for a local model.

The constraint (\ref{dim5prot}) is very strong. Within an F-theory context the question is can it be weakened through factors coming from Yukawa couplings for example. In a minimal $SU(5)$ GUT context the operator is induced through the heavy Higgs triplets. In that case there is a strong suppression due to Yukawa couplings which relaxes somewhat the constraint (\ref{dim5prot}). In the case of F-theory models where all the generations come from a single matter curve it is not clear what the suppression is in going to the lighter generations since it relies on evaluating KK mode wavefunction overlaps. We return to this later in the section. Possible suppression can occur by separating the matter curves so that there is a geometric wavefunction suppression \cite{09052289} though this does not apply to models based on a point of $E_8$ enhancement and is also limited by the finite size of $S$. See also \cite{Leontaris:2009wi} for studies of suppressing proton decay in F-theory by raising the unification scale.

In the case where the generations come from 3 different matter curves the $U(1)$ charges can be used to suppress the dimension 5 proton decay operator for lighter generations. In this case it is also possible to calculate this suppression since it is given exactly by the CKM matrix. Requiring a realistic CKM matrix implies that for the $\te$ matter the top generation is suppressed compared to the charm generation by a factor of $\epsilon^2$ and the top generation is suppressed compared to the up generation by a factor of $\epsilon^3$ where $\epsilon$ is the Wolfenstein parameter $\epsilon \sim 0.2$. In the lepton sector a suppression factor in going from a $\tau$ to a $\mu$ gives an extra factor of $\epsilon^2$. This implies that for a dimension 5 operator involving only the heaviest generations we expect a suppression of $\epsilon^{10} \sim 10^{-7}$. Therefore we estimate that $W_{ijkl}<\frac{10^{-4}}{M_{GUT}}$ for any generations. This seems a little strong compared with minimal $SU(5)$ models which are usually a couple of orders of magnitude weaker. However these are usually evaluated at very small $\mathrm{tan}\beta$ while in F-theory GUTs since the bottom Yukawa appears on the same footing as the top Yukawas we expect large $\mathrm{tan}\beta$\footnote{Note that the amplitude goes like $\left(\mathrm{tan}\beta\right)^2$ and so going from small to large tan$\beta$ can give an enhancement of $10^2-10^3$.}. There are other factors present and we refer to \cite{Murayama:2001ur,Raby:2002wc,Senjanovic:2009kr} for more discussions.

\subsubsection*{Inducing proton decay operators}

The low energy effective theory in a semi-local model comes from an 8-dimensional $E_8$ gauge theory. The theory is then compactified and Higgsed to obtain a 4-dimensional GUT theory. This means that from the cubic interaction term of the 8-dimensional theory we obtain in the 4-dimensional theory the following superpotential interactions
\be
W \supset \f \te \te + \fb \fb \te + X \f \fb + X \te \teb + X X X \;.  \label{cubicwint}
\ee
The massless modes that participate in these interactions can be rendered chiral by an appropriate $U(1)$ flux. On top of these we also have a tower of KK modes with the same gauge charges but which are non chiral and have a mass coming from the profile in the internal directions. We can model these by adding an effective superpotential operator
\be
W \supset M \f^{KK} \fb^{KK} + M \te^{KK} \teb^{KK} \;, \label{kkmassesw}
\ee
here $M$ is a mass parameter which we come back to in more detail at the end of this section but for now it is sufficient to note it is of order $M_{GUT}$. We also need to add KK versions of the operators in (\ref{cubicwint}) where each field may be taken with a KK index.

The operators of (\ref{cubicwint}) and (\ref{kkmassesw}) all come from the interaction in the 8-dimensional theory
\be
W^{8D} \supset \Phi^c \bar{\partial}_{A} \Phi \;,
\ee
where $\Phi$ and $\Phi^c$ denote (different) 8-dimensional fields. The field $\Phi$ is composed of a massless mode $\Phi^0$ and massive modes $\Phi^i$
\be
\Phi = \Phi^0 + \sum_i \Phi^i \;, \;\; \bar{\partial}_A \Phi^0 = 0 \;,\;\; \bar{\partial}_A \Phi^i = M_i \Phi^i \;.
\ee
The KK masses are apparent, while the cubic interactions are given by fluctuations in $A$.

Finally we also have potential masses for the zero modes coming from vevs for $E_8$ singlets such as moduli. The vev for such singlets is not determined but we expect it to be around the GUT scale. Therefore if a mass term is gauge invariant under the full $E_8$ we assume it is present and large
\be
W \supset \left<\phi\right> \f \fb + \left<\phi\right> \te \teb\;,
\label{masswint}
\ee
where $\phi$ denotes a generic $E_8$ singlet.

It is important to note that not all the parameters of (\ref{cubicwint}) and (\ref{masswint}) must be present with order 1 coefficients. Each interaction is multiplied by an integral over the internal dimensions and so may be suppressed or even forbidden if an appropriate geometric symmetry is present.

In minimal GUTs proton decay operators are induced by integrating out the Higgs triplets in the $\f$ representations. However this assumes a coupling between the heavy triplets. In F-theory models there are two possible approaches to higher dimension operators. The first is to allow all possible operators unless constrained by some symmetries of the theory. This accounts for the presence of heavy string modes that might not be accounted for by the renormalisable couplings of the low energy effective theory. The second approach is to consider a higher dimension operator present only if it is induced from renormalisable couplings by integrating out heavy  modes that are present in the theory. We label the first approach stringy and the latter field theoretic.

\noindent
{\bf The field theoretic approach to higher dimension operators}

\noindent
To see how operators are induced in this approach it is useful to consider an example model. We take the model of \cite{Marsano:2009wr} which is based on a 3+2 monodromy group. The matter content of the model is shown in table \ref{tab:model32}.
\begin{table}[ht]
\center
{\scriptsize
\begin{tabular}{|ccccc|}
\hline
Field & Curve & $N_{Y}$ & $M_{U(1)}$ & Exotics  \\
\hline
$\f_{H_u}$   & $-2t_2$    &  +1  & 0  &   \\
$\fb_{H_d}$  & $t_1+t_2$  &  -1  & 0  &   \\
$\fb_{M}$    & $2t_1$     &  0   & -3 &   \\
$\te_{M}$    & $t_2$      &  +1  & +4 & $(3,2)_{+1/6} + 2\times(1,1)_{+1}$  \\
$\teb_{2}$   & $-t_1$     &  -1  & -1 & $(\bar{3},2)_{-1/6} + 2\times(1,1)_{-1}$  \\
\hline \hline
Singlet & Curve & vev & &  \\
\hline
$X_1$ & $t_1-t_2$ & $\epsilon_1$ &  &  \\
$N$ & $-t_1+t_2$ &  &  & Right-handed neutrino  \\
\hline \hline
Induced mass & \multicolumn{4}{c|}{Exotics lifted}  \\
\hline
$\epsilon_1 \te_M \teb_2$ & \multicolumn{4}{c|}{$(\bar{3},2)_{-1/6}(3,2)_{+1/6} + 2\times(1,1)_{-1}(1,1)_{+1}$} \\
\hline \hline
Operator & Charges & Super/Kahler potential & \multicolumn{2}{c|}{Induced?} \\
\hline
$\te_{M}\te_{M}\te_{M}\fb_{M}$ & $2t_1+3t_2$ & W & \multicolumn{2}{c|}{$\epsilon_1\te_{M}\te_{M}\te_{M}\fb_{M}$} \\
$\fb_{M}\fb_{M}\te_M$ & $4t_1+t_2$ & W & \multicolumn{2}{c|}{} \\
$\f_{H_u}\fb_{H_d}$ & $t_1-t_2$ & W & \multicolumn{2}{c|}{} \\
$\f_{H_u}\fb_{M}$ & $2t_1-2t_2$ & W & \multicolumn{2}{c|}{}  \\
\hline
\end{tabular}
}
\caption{Table showing flux restrictions, induced exotics, singlet vevs and induced operators for a model based on a $3+2$ splitting \cite{Marsano:2009wr}.}
\label{tab:model32}
\end{table}
Table \ref{tab:model32} is split into 4 sections. The first shows the curves, their charges, flux restrictions and matter content. The second section shows the GUT singlets and whether they develop a vev or not. The third section shows which exotics are lifted by the appropriate vev. The last section shows if some of the dangerous operators are induced by the singlet vev. We see that a dimension 5 proton decay operator is induced by the singlet vev. It arises from integrating out the KK states in the interactions
\be
W \supset X_1 \f_M^{KK} \fb_{H_d}^{KK} + \fb_{H_d}^{KK}\fb_M \te_M + \f^{KK}_{H_u} \te_M \te_M + X_1 \fb_{H_u}^{KK} \f_{H_d}^{KK} \;, \label{32prde}
\ee
where here and henceforth we drop the dimensionful masses $M$ and also the KK masses $M \f \fb$.

It is important to note that since the Yukawa type interactions involve the KK states, for the case where all the generations come from a single curve 
they do not behave like the Yukawa couplings. This is because the KK wavefunctions are not holomorphic and so the mechanism proposed in \cite{Heckman:2008qa} for the Yukawa couplings is altered. This could mean that the proton decay operator for heavier generations is not as strongly suppressed compared to the lighter ones as would be the case if the Yukawa couplings were used, though we leave a more thorough study for future work.

The constraints on proton decay therefore imply the vev $\epsilon_1$ can not be too large. This in turn implies the exotics can not obtain a large mass and so can affect strongly gauge coupling unification. However a definite statement on how large $\epsilon_1$ can be is difficult to make given the model dependent parameters.

It is worth recalling some facts about gauge coupling unification for the MSSM. See \cite{Amsler} for a review. At 1-loop the couplings unify to around 0.5 percent accuracy. However at 2-loops threshold effects at the TEV scale induce corrections which in turn must be canceled by threshold effects at the GUT scale. These latter effects must be of the order of around 3 percent. Therefore when the exotics are included we require that the gauge couplings unify to around 3 percent at 1-loop to be compatible with MSSM unification. We do not consider the effects of the exotics at 2-loops.

With this in mind we can return to the exotics and note that if we place the exotics spectrum of table \ref{tab:model32} at a scale $2\times10^{12}$GEV the gauge couplings read at the GUT scale, at 1-loop, $\alpha^{-1}_1=20.5$, $\alpha^{-1}_2=19.9$ and $\alpha^{-1}_3=21.2$, which unifies to an accuracy of 6 percent. If we take them down to $2\times10^{10}$GEV unification is at 12 percent.

A possible way out of this tension is to consider additional $U(1)$ symmetries and attempt to lift all the exotics while still preventing proton decay. This is the main theme of this paper.

\noindent
{\bf The stringy approach to higher dimension operators}

\noindent
Before proceeding we return to the alternative approach to higher dimension operators which is to allow for them unless forbidden by symmetries. Here there is not much to say and the preceding analysis is not required. However there is a subtlety to this approach concerning inverse powers of fields. Once the singlets gain a vev in this approach we should write down all operators that involve positive powers of the vevs and are gauge invariant. However this misses out on the possibility that a vev appears with an inverse power and so an opposite $U(1)$ charge. This can occur if the singlet vev gives a mass to an otherwise massless field as is the case with the lifted exotics. Integrating out this field can induce operators with inverse powers of the vevs. The divergence as the vev goes to zero simply signals having integrated out a massless field. In these cases since the integrated out fields must be massless before the singlet vev the field theoretic approach of only allowing such an operator if it comes from integrating out a field/operator already present in the theory is the correct one. We give more explicit examples of this in the model building sections.
	
To summarise, in the stringy approach, we allow for any operators with positive powers of the singlet vevs that are not forbidden by symmetries to be present but only allow for operators involving negative powers of singlet vevs if we can identify the corresponding mode and operators that are integrated out.

\subsubsection*{Exotics and multiple $U(1)$ models}

In section \ref{sec:semloc} we calculated how the hypercharge flux restricts to the matter curves for monodromy groups which allow multiple $U(1)$ symmetries. We can use this to make some general statements about the possibility of lifting all the exotics while preserving a $U(1)$ symmetry to protect against proton decay. First we note that given that the Yukawa couplings must be neutral under all $U(1)$s the charge of the dimension 5 proton decay operator is opposite to that of the $\mu$-term. Therefore the symmetry which forbids proton decay should be a Pecci-Quinn symmetry in that at least one of the Higgs curves must be charged under it. Next we note the following property of the hypercharge restriction to the curves: given a $t_i$ factor, say $t_1$, if we sum the hypercharge restriction over all the $\f$ curves  weighted by their charge under $t_1$ this is equal to minus the sum over the $\te$ curves weighted by the same $t_1$ charge.\footnote{This corresponds to the sum of the hypercharge flux vanishing on each factor of the spectral cover.} This means that to have no exotics charged under a symmetry we require that the sum over the $\f$ curves weighted by the charges under that symmetry vanishes. This in turn implies that the Higgs curves can not be charged under that symmetry since either they have non-trivial hypercharge restrictions in which case there must be exotics on one of the other curves charged under that symmetry, or there is vanishing hypercharge restriction to the Higgs curves in which case there are triplet exotics on the curves which are charged under the symmetry. Hence such a conserved symmetry can not be a PQ symmetry.

Practically we seem to find that the above conclusion indeed holds and there are no models in which all the exotics are lifted and a $U(1)$ remains unbroken to protect against proton decay. The fact that the $U(1)$s are broken is a necessary but not sufficient condition for a model to induce proton decay operators. However practically we find that in all the models we could construct, if we allow for all the interactions allowed by the $U(1)$ charges selection rules, proton decay operators are always induced.\footnote{In all the models where the generations come from a single matter curve the proton decay operators are forbidden by the $U(1)$ charges as long as only positive powers of singlet vevs are allowed. However we find that integrating out the exotics always generates proton decay operators with inverse powers of singlet vevs.}\footnote{Some of the models we find offer the possibility of avoiding proton decay by simply taking the exotics to be very light. It turns out that this still maintains gauge coupling unification because of the particular exotics spectrum that is induced (see section \ref{sec:models2111} for example). However we find that the models which allow for this possibility also require an extra selection principle to avoid an operator of the form $\f_{H_u}\fb_M$ which leads to large neutrino masses.}
As a result our models have to utilise some extra selection rule which we choose to be an R-parity imposed by hand on the curves. It is important to emphasise that it may be possible to utilise some other selection rules which could follow from, for example, geometric separation of the matter curves.

\subsubsection*{Physical and holomorphic exotics mass}

Here we discuss a small subtlety to do with physical versus holomorphic operators. The suppression scales discussed are for physical operators meaning that the fields have canonical kinetic terms. Also the masses of the exotics should be canonically normalised. We now briefly show that the naive intuition is correct at least with respect to scaling with the overall CY base volume. In the superpotential the higher dimension operators can only be suppressed by the Planck scale since the string scale can not appear due to holomorphy which means that for dimension 5 proton decay operator induced by a singlet vev we have the physical suppression \cite{Conlon:2007zza}
\be
{\cal L}_{\mathrm{phys}} = \left(\frac{<X>}{\sqrt{Z_X}M_p} \right) \frac{\te \te \te \f}{Z_M^2M_pe^{-K/2}} \equiv \epsilon_X \frac{\te \te \te \f}{M_{\mathrm{GUT}}} \;.
\ee
Here $K$ is the closed string Kahler potential which has a factor of $-2\ln\;{\cal V}$. $Z_X$ is the kinetic normalisation factor of the singlet field which is not important as long as the vev is a free parameter, i.e. all the physics will concern $\epsilon_X$. $Z_M$ is the matter kinetic normalisation which goes as $Z_M \sim {\cal V}^{-2/3}$ \cite{Conlon:2006tj}. Now consider the mass term for the exotics
\be
{\cal L}_{\mathrm{phys}} = \epsilon_X \frac{e^{K/2}M_p \te \teb}{Z_M} = \epsilon_X M_{\mathrm{GUT}} \te \teb \;.
\ee
Hence we see that indeed the singlet vev sets the mass for the exotics with respect to the GUT scale.

\subsection{R-parity}
\label{sec:rparity}

As discussed above, in all our models we find that the $U(1)$ symmetries are not sufficient to forbid dimension 5 proton decay. We have to forbid these by hand and the primary motivation is the possibility of extra discrete symmetries. The leading candidate is (an extended version of) R-parity (or matter parity) where the Higgs fields are assigned positive parity and the matter fields are assigned negative parity. The GUT singlets and curves holding exotic fields can assigned positive or negative parity.

The R-parity assignment has different meaning according to whether we are adopting the field theoretic or stringy approach to higher dimensional operators as in section \ref{sec:exprot}. In the stringy approach the R-parity is assigned to only the massless modes and all operators involving the massless modes that are allowed by the $U(1)$ symmetries and R-parity are induced. This is the mild version of R-parity.

In the field theoretic approach R-parity is assigned to a curve and also the high mass KK (and string) modes on that curve have that R-parity charge. This implies that a higher dimensional operator that is allowed by R-parity could still not be present if the renormalisable operators that generate it once the KK modes are integrated out are forbidden by R-parity. This latter use is a strong version of R-parity.

Indeed it is possible to show that in this approach an R-parity assignment where all the non singlet curves apart from the Higgs curves are assigned negative charge and all the singlet curves are assigned positive charge, when combined with the $U(1)$ symmetries, always forbids proton decay as long as the only operators present are those discussed in section \ref{sec:exprot}. To see this note that since all the exotics are assigned negative R-parity the only possible cubic couplings involving matter curves are of the type
\be
W \supset \overline{\tilde{\f}}_{H_d} \fb \te + \tilde{\f}_{H_u} \te \te \;.
\ee
Here $\tilde{\f}_{H_u}$ can be either $\f_{H_u}$ or $\f_{H_u}^{KK}$ and the $\f$ and $\te$ without an H subscript denote curves holding either MSSM matter or exotic matter. R-parity by itself also allows for operators such as $\f_{H_d}^{KK} \te \te$ but these are not allowed by the $U(1)$ symmetries since $H_u$ and $H_d$ must have different charges. Proton decay must be induced starting from these operators and integrating out other heavy states. However the net coupling at the end must couple $\overline{\tilde{\f}}_{H_d} \tilde{\f}_{H_u}$. Such a coupling can only be induced by integrating out heavy states starting from the operators
\be
W \supset X \overline{\tilde{\f}}_{H_d} \f + X \tilde{\f}_{H_u} \fb \;, \label{pre3}
\ee
where $X$ stand for some GUT singlets. Note that the vector partner in each coupling can not be a Higgs curve as long as a $\mu$-term is forbidden by the $U(1)$ symmetries. Then we see that the terms (\ref{pre3}) are forbidden by R-parity. Hence we conclude that such an R-parity forbids proton decay operators. This result can be applied to any models where the Higgs curves do not have any massless exotics. In other cases the coupling needs to be studied on a case-by-case basis using the $U(1)$ symmetries.

Having discussed the benefits of the introduced R-parity it is important to emphasise that the origin of such a symmetry is not clear and would also require a global completion to realise. For these reasons we regard the introduction of this symmetry as the weakest phenomenological aspect of the models presented. We refer to \cite{Hayashi:2009bt} for some initial attempts at finding such a symmetry within an F-theory context.

Finally we note that this symmetry could be replaced by some other selection principle such as a geometric separation of curves leading to wavefunction overlap suppression of some operators.

\subsection{Neutrinos}

The neutrino sector is quite model dependent. However there are some general comments that can still be made. Recall that there are two phenomenologically appealing neutrino scenarios studied in F-theory GUTs: the Dirac and Majorana scenarios \cite{ArkaniHamed:2000bq,Bouchard:2009bu}. In the case of a Dirac scenario superpotential Dirac and Majorana masses are forbidden while the Neutrinos obtain a Dirac mass from the Kahler potential. In the Majorana scenario there are both Dirac and Majorana neutrino masses in the superpotential.

We find realisations of the Dirac scenario in most of the models. However there is a tension with the $\mu$-term generation through the Giudice-Masiero mechanism. The problem is that we find that giving an F-term to an appropriate singlet to generate an $\mu$-term also implies that Majorana masses are generated for the right-handed neutrinos through operators in the Kahler potential which are generically at the TEV scale. Since in the Dirac scenario the neutrinos already have eV scale masses the extra suppressions from the Majorana masses makes them too light. We refer to section \ref{sec:models2111} for explicit examples of these issues (and where we also present a resolution to this problem by making the exotics very light).

The Majorana scenario can also be realised. There are two possibilities for right-handed neutrino candidates. The first is for them to be singlets under not only the GUT group but also the extra $U(1)$s \cite{Bouchard:2009bu}. Then a Majorana mass is naturally expected. However there is then the following problem: a superpotential Dirac mass $\f_{H_u} \fb_M N$ implies that also the R-parity violating term $\f_{H_u} \fb_M$ is allowed which in turn must be forbidden by hand. The case where the right-handed neutrinos are taken completely neutral is not model dependent in that it can work for any model and so we have nothing new to say on the matter and do not consider this option in any further detail. There is another Majorana option, which is the one we explicitly realise, which is to take the right-handed neutrinos to be GUT singlets but charged under the $U(1)$s. The Majorana mass can then be generated once some singlets develop a vev \cite{Dudas:2009hu}.

With both neutrino scenarios we find that there is a always a linear term induced by the singlet vevs. This requires the use of an additional R-parity assignment to forbid.

Finally we note that for the single $U(1)$ model the neutrino scenario is problematic since there is no superpotential Dirac mass which means only the Kahler potential Dirac scenario is available. However the singlet also induces a Majorana mass $\epsilon_1^2NN$ which suppresses the neutrino masses too much.

\section{Single curve models}
\label{sec:singcur}

In this section we present some models, based on the setup where all 3 of the SM generations come from a single matter curve. We allow ourselves to choose freely the flux restriction $M$ and $N$ parameters in tables \ref{tab:curves2111}, \ref{tab:curves221} and \ref{tab:curves311}. With these specified the spectrum is determined. The operators are then determined by the $U(1)$ symmetries and by an imposed R-parity assignment. We study the resulting phenomenology and the relation to the constraints discussed in section \ref{sec:phenocon}. In all the models we only use the mild version of R-parity which is just imposed on the massless modes (see section \ref{sec:rparity}). Models where the strong version of R-parity is used and where automatically proton decay is therefore forbidden are studied in the appendix.

\subsection{Models from $2+1+1+1$ factorisation}
\label{sec:models2111}

The model building for this monodromy splitting amounts to assigning the appropriate curves to each matter representation and specifying the flux parameters in table \ref{tab:curves2111}. Further we can give a vev to the available charged singlets which for this monodromy are (for the full orbits see \cite{Dudas:2009hu})
\bea
\un_1 &:& \pm\left(t_1-t_3\right) \;, \nn \\
\un_2 &:& \pm\left(t_1-t_4\right) \;, \nn \\
\un_3 &:& \pm\left(t_1-t_5\right) \;, \nn \\
\un_4 &:& \pm\left(t_3-t_4\right) \;, \nn \\
\un_5 &:& \pm\left(t_3-t_5\right) \;, \nn \\
\un_6 &:& \pm\left(t_4-t_5\right) \;.
\eea
The curves for the up Higgs and $\te$ matter representations are fixed to be
\bea
\f_{H_u} &=& -2t_1 \;, \nn \\
\te_{M} &=& t_1 \;.
\eea
Note that we have identified $t_1 \leftrightarrow t_2$ and denoted both as $t_1$, a notation that we use henceforth. Next we specify the Higgs down curve. There are two choices for this curve: either it involves $t_1$ or it does not. The former case allows for the Giudice-Masiero interaction $X^{\dagger}\f_{H_u}\fb_{H_d}$ in the Kahler potential while the latter case does not. However, the former choice also implies that proton decay operators are generated by the singlets that are required to obtain a vev to lift the exotics. We give an example in the appendix. Therefore we choose the down Higgs curve to not have a $t_1$ factor which without loss of generality implies
\be
\fb_{H_d} = t_3+t_5\;.
\ee
The Giudice-Masiero operator will in turn be generated once the singlet vevs are accounted for. Given this choice the $\f$-matter curve is determined by the requirement of a renormalisable bottom Yukawa to be
\be
\fb_{M} = t_1 + t_4 \;.
\ee
We note that phenomenologically it is not unreasonable to induce the bottom Yukawa coupling through a non-renormalisable operator involving a singlet vev since the bottom quark is much lighter than the top. However we find that in our models the appropriate singlet also led to a $\mu$ term being generated once combined with the other singlets that develop a vev. Therefore we avoid this option.

With the appropriate curves specified it remains to specify the restriction of the fluxes. The first requirement is to induce doublet-triplet splitting on the Higgs curves. This can be achieved in two ways. The first well-known way is by having a non-trivial restriction of the hypercharge flux to the Higgs curves \cite{Beasley:2008kw}. There is a second possibility which is to induce doublet-triplet splitting on some other $\f$-matter curve such that the Higgs triplets can pair up with the exotic triplets and gain a mass once the appropriate GUT singlets develop a vev. We find that the latter method leads to more phenomenologically attractive models and so we study this in this section. Models of the former type
are discussed in the appendix.

The model is based on a $2+1+1+1$ monodromy group and is shown in table \ref{tab:modelleastexot2111}. It has hypercharge flux choices $N_7=-1$, $N_8=+1$, $N_9=0$.
\begin{table}[ht]
\center
{\scriptsize
\begin{tabular}{|cccccc|}
\hline
Field & Curve & $N_{Y}$ & $M_{U(1)}$ & Exotics & R-parity \\
\hline
$\f_{H_u}$   & $-2t_1$ &  0 & +1  & $(3,1)_{-1/3}$ & + \\
$\fb_{H_d}$  & $t_3+t_5$  &  -1 & 0 &  & + \\
$\fb_{M}$    & $t_1+t_4$  &  0 & -3 &  & - \\
$\f_{1}$     & $-t_1-t_3$ &  0 & 0 &  &  \\
$\fb_{3}$    & $t_1+t_5$  &  0 & 0 &  &  \\
$\f_{4}$     & $-t_3-t_4$ &  0 & 0 &  &  \\
$\fb_{6}$    & $t_4+t_5$  &  +1 & -1 & $(\bar{3},1)_{+1/3}$ & - \\
$\te_{M}$    & $t_1$      &  0 & +3 &  & - \\
$\teb_{2}$    & $-t_3$     &  -1 & -1 & $(\bar{3},2)_{-1/6} + 2\times(1,1)_{-1}$ & - \\
$\te_{3}$    & $t_4$      &  +1 & +1 & $(3,2)_{+1/6} + 2\times(1,1)_{+1}$ & - \\
$\te_{4}$    & $t_5$      &  0 & 0 &  &  \\
\hline \hline
Singlet & Curve & vev & F-term & & R-parity \\
\hline
$X_1$ & $t_1-t_4$ & $\epsilon_1$ &  &  & + \\
$X_2$ & $t_1-t_5$ & $\epsilon_2$ &  &  & - \\
$X_3$ & $t_3-t_1$ & $\epsilon_3$ & $\left<F_3\right>$ &  & + \\
$X_4$ & $t_3-t_5$ & $\epsilon_1$ &  &  & + \\
\hline \hline
Induced mass & \multicolumn{4}{c}{Exotics lifted} & R-parity\\
\hline
$\epsilon_1 \f_{H_u} \fb_M$ & \multicolumn{4}{c}{} & - \\
$\epsilon_1 \epsilon_2 \f_{H_u} \fb_6$ & \multicolumn{4}{c}{$(3,1)_{-1/3}(\bar{3},1)_{+1/3}$} & + \\
$\epsilon_1 \epsilon_3 \te_3 \teb_2$ & \multicolumn{4}{c}{$(\bar{3},2)_{-1/6}(3,2)_{+1/6} + 2\times(1,1)_{-1}(1,1)_{+1}$} & + \\
$\epsilon_3 \te_M \teb_2$ & \multicolumn{4}{c}{} & + \\
\hline \hline
Operator & Charges & Super/Kahler potential & \multicolumn{2}{c}{$U(1)$ Neutrality} & R-parity\\
\hline
$\f_{H_u}\fb_{H_d}$ & $-2t_1 +t_3+t_5$ & W & \multicolumn{2}{c}{} & + \\
$\fb_{M}\fb_{M}\te_M$ & $3t_1+2t_4$ & W & \multicolumn{2}{c}{} & - \\
$\te_{M}\te_{M}\te_{M}\fb_{M}$ & $4t_1+t_4$ & W & \multicolumn{2}{c}{} & + \\
$\f_{H_u}\fb_{M}$ & $-t_1+t_4$ & W & \multicolumn{2}{c}{$\epsilon_1\f_{H_u}\fb_{M}$} & - \\
$\left<F^{\dagger}\right>\f_{H_u}\fb_{H_d}$ & $-2t_1 +t_3+t_5$ & K & \multicolumn{2}{c}{$\epsilon_3\epsilon_4\left<F^{\dagger}_3\right>\f_{H_u}\fb_{H_d}$} & + \\
\hline
\end{tabular}
}
\caption{Table showing flux restrictions, induced exotics, singlet vevs and induced operators with positive powers of singlet insertions for a model based on a $2+1+1+1$ splitting.}
\label{tab:modelleastexot2111}
\end{table}

Table \ref{tab:modelleastexot2111} is split into 4 sections. The first lists the matter curve assignments, the flux restrictions, the exotics spectrum and the assigned R-parity charges. The second section lists the GUT singlets that have a vev, possible F-terms and the assigned R-parity charges. The third section shows the induced mass for the exotics and which exotics are lifted by which singlet vevs. The final column shows the R-parity charge of the full mass operator. The fourth section shows whether other important operators are allowed by the $U(1)$ symmetries with only positive powers of singlet insertions and by R-parity.

Consider first gauge coupling unification. If we give the singlets vevs $\epsilon_1\epsilon_2=\epsilon_1\epsilon_3=2\times10^{15}$GEV the gauge couplings at the GUT scale read $\alpha^{-1}_1=23.2$, $\alpha^{-1}_2=23.2$ and $\alpha^{-1}_3=23.1$. This is equivalent to 1-loop MSSM unification. The reason for this is that although the exotics do not form complete GUT representations as far as the beta functions are concerned they act as a complete vector pair of $\te$ representations
\be
 \left[\te\right] = \left[ (3,2)_{+1/6} + (\bar{3},1)_{-2/3} + (1,1)_{+1} \right] \
 \sim \ \left[ (3,2)_{+1/6} + (\bar{3},1)_{+1/3} + 2\times(1,1)_{+1} \right] \;. \label{specexot2}
\ee
Indeed this means that they can be taken all the way down to the TeV scale by taking $\epsilon_1\epsilon_2\sim\epsilon_3 \sim 10^{-13}$ which still gives $\alpha^{-1}_1=10.0$, $\alpha^{-1}_2=10.0$ and $\alpha^{-1}_3=9.9$ at the GUT scale. We find that such combinations of exotics arise frequently in the models.

Consider now proton decay. As is shown in last section of table \ref{tab:modelleastexot2111} proton decay operators are forbidden with only positive powers of singlet insertions by the $U(1)$ charges. However let us forget about the R-parity assignments and consider the following operators allowed simply by the $U(1)$ symmetries
\be
W \supset \epsilon_1 \f_{H_u} \fb_{M} + \epsilon_1 \epsilon_2 \f_{H_u} \fb_6 + \epsilon_1\epsilon_3 \fb_6 \fb_M \te_M \;. \label{2111wp}
\ee
If we integrate out the exotic pair on $\f_{H_u}$ and $\fb_6$ we generate the effective proton decay operator
\be
W \supset \left(\frac{\epsilon_1\epsilon_3}{\epsilon_2}\right) \fb_M \fb_M \te_M \;. \label{2111badop}
\ee
The pole indicates that as $\epsilon_2 \rightarrow 0$ the exotics become massless. Another way to think of this operator is to realise that the massless mode in (\ref{2111wp}) is a mixture of $\fb_M$ and $\fb_6$ which then induces proton decay from the final term in (\ref{2111wp}). Similar physics occurs by mixing in the $\te$ representations. Once we impose R-parity however the first and last terms in (\ref{2111wp}) are absent and no such operator is generated. It is important to note that the mild version of R-parity (see section \ref{sec:rparity}) which acts only on the massless modes is sufficient for these purposes. This is so even though the final operator (\ref{2111badop}) is not forbidden by any symmetries including R-parity. The reason is that the pole in $\epsilon_2$ could only have come from the terms in (\ref{2111wp}) (and similar ones in the $\te$ sector).

The appropriate R-parity assignments for all the curves are shown in the last column of table \ref{tab:modelleastexot2111}. These amount to forbidding mixing between the matter and exotic representations. They also forbid a direct coupling $\epsilon_1 \epsilon_3\fb_6 \fb_M \te_M$ which would lead to proton decay mediated by the exotic triplets leading to the dimension 5 operator
\be
\left(\frac{\epsilon_3}{\epsilon_2}\right)\fb_M \te_M \te_M \te_M \;.
\ee
Finally we note that they also forbid the dangerous operator $\epsilon_1\f_{H_u}\fb_M$ which would lead to large neutrino masses.

Note that because the exotics spectrum does not affect gauge coupling unification we can consider the possibility of not forbidding proton decay operators but rather suppressing them by taking small singlet vevs and light exotics. In this model this can be realised by taking $\epsilon_1 \sim \epsilon_3 \ll 1$. However the operator $\epsilon_1\f_{H_u}\fb_M$ would still be too large and would require some extra selection rule to forbid. Nonetheless forbidding this operator may be easier to realise than the full R-parity implementation suggested here and so this possibility remains attractive. 

A Giudice-Masiero term in the Kahler potential can lead to a $\mu$-term
\be
K \supset \epsilon_3 \epsilon_4 \left<F_3^{\dagger} \right> \f_{H_u} \fb_{H_d} \;.
\ee
Note that it is for this operator that $\epsilon_4$ must be introduced. A similar term $\epsilon_2\left<F_3^{\dagger} \right>\f_{H_u} \fb_{H_d}$ also satisfies the $U(1)$ constraints but is forbidden by the R-parity assignments.

We now turn to the neutrinos. A Dirac scenario would involve a right-handed neutrino choice of $N=t_3-t_4$ (with positive R-parity charge). The factor of $+t_3$ implies that it can not appear in the superpotential and so there are no superpotential Dirac or Majorana masses which allows for the Kahler potential Dirac scenario. Indeed such Dirac mass terms are obtained through the Kahler potential operator $ \epsilon_3\epsilon_4\f_{H_d}\fb_{M}N$.
However, as discussed in section \ref{sec:phenocon}, in this model and also in the other models considered in this work the Dirac neutrino scenario, where a neutrino mass is generated through the Higgs F-term in the Kahler potential as originally proposed in \cite{ArkaniHamed:2000bq,Bouchard:2009bu}, is in tension with the Giudice-Masiero mechanism. The reason is that the Giudice-Masiero F-term generically induces TeV scale Majorana masses for the right-handed neutrinos. Explicitly we also have the Kahler potential operator $\epsilon_1^2\epsilon_3 \left<F^{\dagger}_3\right> N N$. However in this model because of the exotic spectrum we can take $\epsilon_1$ and $\epsilon_3$ very small to suppress these masses which leads to an interesting connection between the exotics and neutrino masses.

It is also possible to realise a Majorana scenario with charged GUT singlets. For example taking the right-handed neutrino curve to be $N=t_4-t_3$. The resulting neutrino masses are given by the operators
\be
\epsilon_1^2\epsilon_3 \f_{H_u} \fb_M N + \epsilon_1^2\epsilon_3^2 NN \;,
\ee
which give the mass scale
\be
M_{\nu} \sim  \epsilon_1^2 10^{-3} \mathrm{eV} \;.
\ee
This mass is quite light though even with a large vev for $\epsilon_1$.

\subsection{Models from $2+2+1$ factorisation}
\label{sec:models221}

For this monodromy configuration the available charged singlets are (for the full orbits see \cite{Dudas:2009hu})
\bea
\un_1 &:& \pm\left(t_1-t_3\right) \;, \nn \\
\un_2 &:& \pm\left(t_1-t_5\right) \;, \nn \\
\un_3 &:& \pm\left(t_3-t_5\right) \;.
\eea
There are 4 choices for the $\fb_{H_d}$ and $\fb_{M}$ curves. 3 of these have a factor of $+t_1$ in $\fb_{H_d}$ which means that the $\mu$-term can be induced by just one singlet vev. This type of set up always leads to a $\mu$-term or proton decay once the exotics are lifted by the singlets.\footnote{We study one of these possibilities in the appendix where although proton decay operators are not forbidden by the $U(1)$ symmetries they can be forbidden by the strong version of R-parity.} The remaining possibility is
\bea
\f_{H_u} &=& -2t_1 \;, \nn \\
\fb_{H_d} &=& \fb_3 = t_3+t_5 \nn \\
\fb_M &=& \fb_1 = t_1 + t_3 \;.
\eea
The model is shown in table \ref{tab:model221}. It has hypercharge flux choices $N_1=0$ and $N_2=+1$.
\begin{table}[ht]
\center
{\scriptsize
\begin{tabular}{|cccccc|}
\hline
Field & Curve & $N_{Y}$ & $M_{U(1)}$ & Exotics & R-parity \\
\hline
$\f_{H_u}$   & $-2t_1$    &  0   & +1 & $(3,1)_{-1/3}$ & + \\
$\fb_{H_d}$  & $t_3+t_5$  &  -1  & 0  &  & + \\
$\fb_{M}$    & $t_1+t_3$  &  0   & -3 &  & - \\
$\fb_{2}$    & $t_1+t_5$  &  0   & 0  &  &  \\
$\fb_{4}$    &  $2t_3$    &  +1  & -1 & $(\bar{3},1)_{+1/3}$ & + \\
$\te_{M}$    & $t_1$      &  0   & +3 &  & - \\
$\te_{2}$    & $t_3$      &  +1  & +1 & $(3,2)_{+1/6} + 2\times(1,1)_{+1}$ & + \\
$\teb_{3}$   & $-t_5$     &  -1  & -1 & $(\bar{3},2)_{-1/6} + 2\times(1,1)_{-1}$ & - \\
\hline \hline
Singlet & Curve & vev & F-term & & R-parity \\
\hline
$X_1$ & $t_1-t_3$ & $\epsilon_1$ & & & +  \\
$X_2$ & $-t_3+t_5$ & $\epsilon_2$ &  & & -  \\
$X_3$ & $-t_1+t_5$ & $\epsilon_3$ & $\left<F_3\right>$ & & +  \\
\hline \hline
Induced mass & \multicolumn{4}{c}{Exotics lifted} & R-parity\\
\hline
$\epsilon_1 \f_{H_u} \fb_M $ & \multicolumn{4}{c}{} & - \\
$\epsilon_1^2 \f_{H_u} \fb_4$ & \multicolumn{4}{c}{$(3,1)_{-1/3}(\bar{3},1)_{+1/3}$} & + \\
$\epsilon_2 \te_2 \teb_3$ & \multicolumn{4}{c}{$(\bar{3},2)_{-1/6}(3,2)_{+1/6} + 2\times(1,1)_{+1}(1,1)_{-1}$} & + \\
$\epsilon_1\epsilon_2 \te_M \teb_3$ & \multicolumn{4}{c}{} & - \\
$\epsilon_1\epsilon_3 \te_2 \teb_3$ & \multicolumn{4}{c}{} & - \\
$\epsilon_1^2\epsilon_3 \te_M \teb_3$ & \multicolumn{4}{c}{} & + \\
\hline \hline
Operator & Charges & Super/Kahler potential & \multicolumn{2}{c}{$U(1)$ Neutrality} & R-parity \\
\hline
$\f_{H_u}\fb_{H_d}$ & $-2t_1 +t_3+t_5$ & W & \multicolumn{2}{c}{} & + \\
$\fb_{M}\fb_{M}\te_M$ & $3t_1+2t_3$ & W & \multicolumn{2}{c}{$\epsilon_3\fb_{M}\fb_{M}\te_M$} & - \\
$\te_{M}\te_{M}\te_{M}\fb_{M}$ & $4t_1+t_3$ & W & \multicolumn{2}{c}{} & + \\
$\f_{H_u}\fb_{M}$ & $-t_1+t_3$ & W & \multicolumn{2}{c}{$\epsilon_1\f_{H_u}\fb_{M}$} & - \\
$\left<F^{\dagger}\right>\f_{H_u}\fb_{H_d}$ & $-2t_1 +t_3+t_5$ & K & \multicolumn{2}{c}{$\epsilon_1\left<F^{\dagger}_3\right>\f_{H_u}\fb_{H_d}$} & + \\
\hline
\end{tabular}
}
\caption{Table showing flux restrictions, induced exotics, singlet vevs and induced operators with positive powers of singlet insertions for a model based on a $2+2+1$ splitting.}
\label{tab:model221}
\end{table}

The gauge coupling unification scenario is the same as that of the model in section \ref{sec:models2111} with equal accuracy to that of the MSSM. The only requirement is that $\epsilon_1^2 \sim \epsilon_2$.

The key to understanding proton decay in this model is the singlet vev $\epsilon_3$. This singlet is introduced purely to generate a $\mu$-term through its F-term. Therefore it is possible to take $\epsilon_3 \rightarrow 0$. In that case we see that proton decay operators with positive powers of singlet vevs are forbidden by the $U(1)$ symmetries. If we also impose the R-parity assignments we see there is no mixing between the matter and exotics through the mass operators and so it is not possible to generate negative powers of the singlet vevs in this way. This means that dimension 4 proton decay is not possible to induce. Dimension 5 proton decay can only come from the operator\footnote{This would be induced directly by integrating out the exotic triplets if the coupling $\epsilon_2 \fb_4 \fb_M \te_M$ was not forbidden by R-parity.}
\be
\left(\frac{\epsilon_2}{\epsilon_1^2}\right) \fb_M \te_M \te_M \te_M \;,
\ee
which is forbidden by the R-parity.

However we can also turn on $\epsilon_3$ and still no proton decay is induced. We see that the dimension 4 proton decay operator in the last section of table \ref{tab:model221} is forbidden by R-parity. Also because of R-parity $\epsilon_3$ only induces mixing in the $\te$ sector. This means that it is not possible to generate proton decay operators with negative powers of $\epsilon_1$ since in the limit $\epsilon_1 \rightarrow 0$ no exotic $\te$ particles become massless. Now the only possible dimension 5 proton decay operator must involve at least one positive power of $\epsilon_3$ (since we know that as $\epsilon_3\rightarrow 0$ no proton decay is induced) which gives
\be
\left(\frac{\epsilon_3}{\epsilon_1}\right) \fb_M \te_M \te_M \te_M \;.
\ee
Since this has a pole in $\epsilon_1$ it can not be induced.

As shown in table \ref{tab:model221} a Giudice-Masiero term in the Kahler potential can lead to a $\mu$-term
\be
K \supset \epsilon_1 \left<F_3^{\dagger} \right> \f_{H_u} \fb_{H_d} \;.
\ee

We now turn to the neutrino scenarios. Interestingly they can both be realised by $N=t_1-t_5$ according to which R-parity charge it is assigned.
If we take $N$ to have positive R-parity we have the superpotential operators\footnote{Note that $N=-t_1+t_3$ with negative R-parity charge also gives a Majorana scenario.}
\be
W \supset \epsilon_2 \f_{H_u} \fb_M N + \epsilon_3^2 N N \;,
\ee
which lead to a standard Majorana scenario. Note that in this case we have the phenomenologically attractive feature that $\epsilon_3$ only appears in the neutrino masses and so can be adjusted to fit phenomenology.

If we take $N$ to have negative R-parity charge and also take $\epsilon_3 \rightarrow 0$ then we have a Dirac scenario with the neutrino masses coming from the operator
\be
K \supset \left(\epsilon_1\epsilon_2\right)^2 \f_{H_d} \fb_M N \;.
\ee
Note that as usual if we use the Giudice-Masiero mechanism a Majorana mass is also induced form the Kahler potential $\left<F_3^{\dagger}\right>\left(\epsilon_1\epsilon_2\right)^3NN$ which suppresses the neutrino masses too much.

\subsection{Models from $3+1+1$ factorisation}
\label{sec:models311}

For these models, if the bottom Yukawa coupling comes from a renormalisable operator, it is possible to show that lifting the exotics always generates a $\mu$-term at the same scale. To see this note that the $\f_{H_u}$ curve has no hypercharge restriction which means it must have exotics, which in the minimal case are the Higgs triplets. In order to generate a renormalisable bottom Yukawa coupling we are forced to take
\bea
\f_{H_u} &=& -2t_1 \;, \nn \\
\fb_{H_d} &=& \fb_1 = t_1+t_4 \nn \\
\fb_M &=& \fb_2 = t_1 + t_5 \;. \label{311curvenorbot}
\eea
This means that since we must forbid the couplings $\f_{H_u}\fb_{H_d}$ and $\f_{H_u}\fb_{M}$ the Higgs triplet must gain a mass by coupling to a vector partner on the $\fb_{3}$ curve. Now $\f_{H_u}\fb_3=-2t_1+t_4+t_5$ which requires two singlets to obtain a vev. Given that the only singlets available are $X_1 = \left(t_1-t_4\right)$, $X_2 = \left(t_1-t_5\right)$ and $X_3 = \pm\left(t_4-t_5\right)$, giving two of them a vev also generates a $\mu$ term.

This problem can be avoided if the bottom Yukawa comes from a non-renormalisable operator. This is in fact phenomenologically slightly preferable as it can explain the lightness of the bottom quark compared to the top quark. The model is presented in table \ref{tab:model311main} and corresponds to taking $N_2=0$ and $N_1=-1$.
\begin{table}[ht]
\center
{\scriptsize
\begin{tabular}{|cccccc|}
\hline
Field & Curve & $N_{Y}$ & $M_{U(1)}$ & Exotics & R-parity \\
\hline
$\f_{H_u}$   & $-2t_1$    &  0   & +1 & $(3,1)_{-1/3}$ & + \\
$\fb_{H_d}$  & $t_4+t_5$  &  -1  & 0  &  & + \\
$\fb_{M}$    & $t_1+t_4$  &  0   & -3 &  & - \\
$\fb_{2}$    & $t_1+t_5$  &  +1  & -1 & $(\bar{3},1)_{+1/3}$ & - \\
$\te_{M}$    & $t_1$      &  +1  & +4 & $(3,2)_{+1/6} + 2\times(1,1)_{+1}$ & - \\
$\teb_{2}$   & $-t_4$     &  -1  & -1 & $(\bar{3},2)_{-1/6} + 2\times(1,1)_{-1}$ & - \\
$\te_{3}$    & $t_5$      &  0   & 0   &  &  \\
\hline \hline
Singlet & Curve & vev & F-term & & R-parity \\
\hline
$X_1$ & $t_1-t_4$ & $\epsilon_1$ &  & & +  \\
$X_2$ & $t_1-t_5$ & $\epsilon_2$ &  & & +  \\
$X_3$ & $-t_1+t_4$ & $\epsilon_3$ &  & Note: $X_3=X_1^{\dagger}$ & +  \\
$X_4$ & $t_4-t_5$ & $\epsilon_4$ &  & & -  \\
\hline \hline
Induced mass & \multicolumn{4}{c}{Exotics lifted} & R-parity \\
\hline
$\epsilon_1\epsilon_4 \f_{H_u} \fb_2$ & \multicolumn{4}{c}{$(3,1)_{-1/3}(\bar{3},1)_{+1/3}$} & + \\
$\epsilon_1 \f_{H_u} \fb_M$ & \multicolumn{4}{c}{} & - \\
$\epsilon_3 \te_M \teb_2$ & \multicolumn{4}{c}{$(\bar{3},2)_{-1/6}(3,2)_{+1/6} + 2\times(1,1)_{-1}(1,1)_{+1}$} & + \\
\hline \hline
Operator & Charges & Super/Kahler potential & \multicolumn{2}{c}{$U(1)$ Neutrality} & R-parity \\
\hline
$\fb_{H_d}\fb_{M}\te_M$ & $2t_1 + 2t_4+t_5$ & W & \multicolumn{2}{c}{$\epsilon_1\fb_{H_d}\fb_{M}\te_M$} & + \\
$\f_{H_u}\fb_{H_d}$ & $-2t_1 +t_4+t_5$ & W & \multicolumn{2}{c}{$\epsilon_1\epsilon_2\f_{H_u}\fb_{H_d}$} & + \\
$\fb_{M}\fb_{M}\te_M$ & $3t_1+2t_4$ & W & \multicolumn{2}{c}{} & - \\
$\te_{M}\te_{M}\te_{M}\fb_{M}$ & $4t_1+t_4$ & W & \multicolumn{2}{c}{} & + \\
$\f_{H_u}\fb_{M}$ & $-t_1+t_4$ & W & \multicolumn{2}{c}{$\epsilon_1\f_{H_u}\fb_{M}$} & - \\
$\fb_{2}\fb_{M}\te_M$ & $3t_1+t_4+t_5$ & W & \multicolumn{2}{c}{$\fb_{2}\fb_{M}\te_M$} & - \\
\hline
\end{tabular}
}
\caption{Table showing flux restrictions, induced exotics, singlet vevs and induced operators with positive powers of singlet insertions for a model based on a $3+1+1$ splitting.}
\label{tab:model311main}
\end{table}

As shown in table \ref{tab:model311main} the bottom Yukawa coupling is induced by a higher dimension operator involving $X_1$. This implies that the vev $\epsilon_1$ is not too small.

Unlike the other models here the $\mu$-term is induced in the superpotential by a singlet vev much like in the NMSSM. This forces us to take $\epsilon_2 \sim 10^{-13}$.

Gauge coupling unification works in exactly the same way as the previous two models since the exotics spectrum is the same: it recreates the accuracy of the MSSM at 1-loop.

We see from table \ref{tab:model311main} that proton decay operators are forbidden with only positive powers of singlet insertions. Also the operator $\fb_2 \fb_M \te_M$, which could induce proton decay directly, is forbidden by R-parity. Finally there is no mixing between the matter and exotics and so no proton decay operators are induced.

We now turn to the neutrinos. It is not possible to realise the Majorana scenario. To see this note that such a scenario requires a superpotential Dirac mass which must involve a cubic interaction $\f_{H_u}\fb N$ which in turn requires that $N$ has a $+t_1$ factor. However all the available such singlets are used to give the exotics a mass.\footnote{It may be possible to let the right handed neutrino have a TeV scale vev.}

It is possible to realise a Dirac scenario though it has a problem. We take $N=-t_1+t_5$ which gives a Kahler potential Dirac mass $\f_{H_d} \fb_M N$. As discussed above there is no Dirac superpotential mass. There is a superpotential Majorana mass but this is of the form $\epsilon_2^2 NN$. Since $\epsilon_1\epsilon_2 \sim 10^{-13}$ the ratios of Dirac to Majorana masses are
\bea
\frac{M_D^2}{M_W} &\sim& \epsilon_1^2 10^{-3} \mathrm{eV} \;, \nn \\
\frac{M_D}{M_W} &\sim& \frac{\epsilon_1^2}{10} \;.
\eea
The problem here is that $\epsilon_1$ must be taken large to get even close to the required neutrino masses. However this then implies that the Majorana masses are of the same order as the Dirac masses which leads to too large disappearance rates into sterile neutrinos.

\section{3-curve models}
\label{sec:mulcur}

In \cite{Dudas:2009hu} a study was initiated of models where each of the SM generations resides on a different matter curve. So 3 $\f$-curves and 3 $\te$-curves for the 3 generations. The motivation for this is to account for the flavour hierarchies in the SM by using the different $U(1)$ charges of the generations through the Froggatt-Nielsen mechanism \cite{Froggatt:1978nt}. As shown in \cite{Dudas:2009hu} such models have to based on a $2+1+1+1$ monodromy splitting. Further there is a unique choice of matter curves that can reproduce realistic Yukawa couplings while avoiding proton decay and a $\mu$-term which are the curves given in table \ref{tab:model21113cur}. Given these matter curves there is some choice as to which singlets develop a vev and which are assigned to right-handed neutrinos. In \cite{Dudas:2009hu} it was shown that the only phenomenologically compatible possibilities for the singlets that develop a vev are given by
\be
X_1 = -t_3+t_4 \;,\;X_2 = t_1-t_4 \;.
\ee
It is possible for more singlets to develop a vev but at least these 2 are required for the quark sector.

For these models we find that a dimension 5 proton decay operator is always allowed by the $U(1)$ symmetries once the exotics are lifted. To see this note that since each generation comes from a different $\te$ curve, if we allow any generation indices the proton decay operator can at most be protected by one $U(1)$ symmetry which we associate to $t_5$. This means that no singlets with a $+t_5$ factor can obtain a large vev. However once a non-trivial hypercharge flux is introduced, on any curve, it must be that there are exotics charged under $t_5$. This can be seen as follows: consider the $\te$ curves, since on the 3 $\te$ matter curves corresponding to the MSSM we must have net positive chirality, the only vector partners to the exotics induced on these curves must come from the final $\te$ curve and charged as $-t_5$. This means that to lift them we require a singlet with a $+t_5$ factor which induces proton decay.

This problem means that it is not possible to forbid proton decay using the mild version of R-parity acting only on the massless modes (see section \ref{sec:rparity}). In this sense it is on a similar footing to the single $U(1)$ model of \cite{Marsano:2009wr}. If we impose the stronger version of R-parity and only consider proton decay operators that are induced from renormalisable operators by integrating out KK modes then proton decay is forbidden due to the result of section \ref{sec:rparity}. We assume this is the case and study the resulting model.

Up to extra pairs of exotics the matter content of the model is unique and is shown in table \ref{tab:model21113cur}. The model has hypercharge flux choices $N_7=-2$, $N_8=+2$ and $N_9=+1$.
\begin{table}[ht]
\center
{\tiny
\begin{tabular}{|cccccc|}
\hline
Field & Curve & $N_{Y}$ & $M_{U(1)}$ & Exotics & R-parity \\
\hline
$\f_{H_u}$   & $-2t_1$ &  +1 & 0  & - & +\\
$\fb_{H_d}$  & $t_3+t_5$  &  -1 & 0  &  & + \\
$\fb_{b}$    & $t_1+t_4$  &  -1 & -1 & $(1,2)_{-1/2}$ & - \\
$\fb_{s}$    & $t_1+t_3$ &  -1 & -2 & $(\bar{3},1)_{+1/3} + 2\times (1,2)_{-1/2}$  & - \\
$\fb_{d}$    & $t_3+t_4$  &  0 & -1 &  & - \\
$\f_{3}$     & $-t_1-t_5$ &  -1 & +1 & $(3,1)_{-1/3}$ & - \\
$\f_{6}$    & $-t_4-t_5$  &  +3 & 0 & $3\times(1,2)_{+1/2}$ & - \\
$\te_{t}$    & $t_1$      &  -1 & +2 & $(3,2)_{+1/6} + 2\times(\bar{3},1)_{-2/3}$ & - \\
$\te_{c}$   & $t_4$      &  +2 & +3 & $2\times(3,2)_{+1/6} + 4\times(1,1)_{+1}$ & - \\
$\te_{u}$    & $t_3$      &  -2 & +3 & $2\times(3,2)_{+1/6} + 4\times(\bar{3},1)_{-2/3}$ & - \\
$\teb_{4}$    & $-t_5$      &  +1 & -5 & $5\times(\bar{3},2)_{-1/6} + 6\times(3,1)_{+2/3} + 4\times(1,1)_{-1}$ & - \\
\hline \hline
Singlet & Curve & vev & F-term & & R-parity \\
\hline
$X_1$ & $-t_3+t_4$ & $\epsilon_1$ &  & & + \\
$X_2$ & $t_1-t_4$ & $\epsilon_2$ &  &  & +\\
$X_3$ & $-t_1+t_5$ & $\epsilon_3$ & $\left<F_3\right>$ & & + \\
\hline \hline
Induced mass & \multicolumn{4}{c}{Exotics lifted} & R-parity\\
\hline
$\epsilon_3 \fb_{b} \f_6$ & \multicolumn{4}{c}{$(1,2)_{-1/2}(1,2)_{+1/2}$} & +\\
$\epsilon_1\epsilon_2\epsilon_3 \fb_{s} \f_3$ & \multicolumn{4}{c}{$(\bar{3},1)_{+1/3}(3,1)_{-1/3}$} & +\\
$\epsilon_1\epsilon_3 \fb_{s} \f_6$ & \multicolumn{4}{c}{$2\times(1,2)_{-1/2}(1,2)_{+1/2}$} & +\\
$\epsilon_3 \te_t \teb_4$ & \multicolumn{4}{c}{$(3,2)_{+1/6}(\bar{3},2)_{-1/6} + 2\times(\bar{3},1)_{-2/3}(3,1)_{+2/3}$} & +\\
$\epsilon_1\epsilon_2\epsilon_3 \te_c \teb_4$ & \multicolumn{4}{c}{$2\times(3,2)_{+1/6}(\bar{3},2)_{-1/6} + 4\times(1,1)_{+1}(1,1)_{-1}$} & +\\
$\epsilon_2\epsilon_3 \te_u \teb_4$ & \multicolumn{4}{c}{$2\times(3,2)_{+1/6}(\bar{3},2)_{-1/6} + 4\times(\bar{3},1)_{-2/3}(3,1)_{+2/3}$} & +\\
\hline \hline
Operator & Charges & Super/Kahler potential & \multicolumn{2}{c}{$U(1)$ neutrality} & R-parity \\
\hline
$\f_{H_u}\fb_{H_d}$ & $-2t_1 +t_3+t_5$ & W & \multicolumn{2}{c}{} & + \\
$\fb_{M}\fb_{M}\te_M$ & ... & W & \multicolumn{2}{c}{...} & - \\
$\te_{M}\te_{M}\te_{M}\fb_{M}$ & ... & W & \multicolumn{2}{c}{...} & + \\
$\f_{H_u}\fb_{b}$ & $-t_1+t_4$ & W & \multicolumn{2}{c}{$\epsilon_2\f_{H_u}\fb_{b}$} & - \\
$\f_{H_u}\fb_{s}$ & $-t_1+t_3$ & W & \multicolumn{2}{c}{$\epsilon_1\epsilon_2\f_{H_u}\fb_{s}$} & - \\
$\f_{H_u}\fb_{d}$ & $-2t_1+t_3+t_4$ & W & \multicolumn{2}{c}{$\epsilon_1\epsilon_2^2\f_{H_u}\fb_{d}$} & - \\
$\left<F^{\dagger}\right>\f_{H_u}\fb_{H_d}$ & $-2t_1 +t_3+t_5$ & K & \multicolumn{2}{c}{$\epsilon_1\epsilon_3\left<F^{\dagger}_3\right>\f_{H_u}\fb_{H_d}$} & + \\
$\fb_{E}\fb_{M} \te_M$ & ... & W & \multicolumn{2}{c}{...} & - \\
\hline
\end{tabular}
}
\caption{Table showing flux restrictions, induced exotics, singlet vevs and induced operators with positive powers of singlet insertions for a 3-curve model based on a $2+1+1+1$ monodromy. The ellipses in the last section of the table denote multiple terms present.}
\label{tab:model21113cur}
\end{table}

Consider gauge coupling unification. The highest mass we can give the exotics is around $2\times10^{14}$GEV since $\epsilon_1\epsilon_2 \sim 10^{-2}$ from the quark masses \cite{Dudas:2009hu}. In that case the gauge couplings read $\alpha^{-1}_1=11.4$, $\alpha^{-1}_2=11.1$ and $\alpha^{-1}_3=11.7$ at 1-loop which is within the required threshold corrections for the MSSM. Note that this is at 1-loop and also does not take into account the small mass splitting between the exotic representations.

Note that we require R-parity to forbid the problematic $\f_{H_u}\fb_M$ coupling which in some of the single curve models was forbidden by the $U(1)$ symmetries. Also note that since the extra singlet $X_3$ has a factor of $t_5$ it does not alter the Yukawa couplings presented in \cite{Dudas:2009hu} and so the quark mass hierarchies are retained.

Finally we turn to the neutrinos scenario. First we note that the scenario presented in \cite{Dudas:2009hu} where the right handed neutrinos came from 3 different curves is not possible here since one of those neutrino curves corresponds a singlet used to lift the exotics. It is still possible to realise the scenarios using one curve for all the right handed neutrinos. The Dirac scenario can be realised by taking $N=-t_4+t_5$. The $+t_5$ factor ensures there is no superpotential Dirac or Majorana mass while in the Kahler potential we have the operators $\epsilon_1\epsilon_2\f_{H_d} \fb_b N$, $\epsilon_2\f_{H_d} \fb_s N$ and $\f_{H_d} \fb_d N$. Again we note there is tension with the Giudice-Masiero operator. The Majorana scenario can also be implemented with $N=t_4-t_5$ but the resulting neutrino masses come out too small to be consistent with phenomenology.

\subsection{Froggatt-Nielsen and quark-lepton mass splitting}
\label{sec:frogeorgi}

As shown in \cite{Dudas:2009hu} 3-curve models can recreate realistic quark and lepton mixing at the GUT level. There is a well known GUT flavour puzzle relating to the quark and lepton masses at the GUT scale. The problem is that at the GUT scale the quark and lepton masses read \cite{Georgi:1979df}\footnote{See \cite{Ross:2007az} for more modern and detailed analysis.}
\be
m_{b} \simeq m_{\tau} \;,\; m_{s} \simeq \frac{1}{3}m_{\mu} \;,\; m_{d} \simeq 3m_{e} \;, \label{gjmass}
\ee
whereas a GUT theory would predict the masses to unify since all the representations fill a single $\f$ GUT multiplet. In \cite{Georgi:1979df} a mechanism to account for this was proposed using higher $SU(5)$ representations which are not available in string theory. In terms of F-theory GUTs, since the mass splitting does not respect the GUT symmetries, we expect it to originate from the hypercharge flux. In this section we show that a natural explanation for the mass splitting pattern can originate from combining the Froggatt-Nielsen mechanism with the hypercharge flux GUT breaking.

The Froggatt-Nielsen based 3-curve models rely on higher dimension operators involving the GUT singlets $X_i$ being generated in the superpotential. These should be induced by integrating out string and KK modes. Taking the approach of section \ref{sec:exprot} the relevant terms in the superpotential are
\be
W \supset \fb_{H_d}\fb^{KK}_{b}\te_{t} + X \fb_{s} \f^{KK}_{b} \;.
\ee
Integrating out the KK modes gives
\be
\frac{X}{M_{KK}} \fb_{H_d} \fb_{s} \te_{t} \;.
\ee
This is the Froggatt-Nielsen mechanism at the GUT level. More precisely there is not one KK mode but rather a tower of them and also there are string modes. Also the physical suppression scale of the operator is not the KK scale but rather the GUT scale which is the winding scale \cite{Conlon:2007zza,Conlon:2009qa}. Nevertheless the essence of the mechanism is as discussed. In terms of Feynman diagrams the higher dimensional operator comes from a diagram with $X$, $\fb_{s}$ and $\fb_{H_d}$, $\te_{t}$ exchanging a $\f$ KK state with a mass insertion in the middle. Note that the KK state is associated to the heavier generation, so that bottom KK modes contribute to strange Yukawas and so on.

Now consider the case, as in table \ref{tab:model21113cur}, where there is non-trivial hypercharge flux restricted to say the $\f_b$ and $\f_s$ curves (this is automatic with a non-trivial restriction to the $\f_{H_u}$ curve). Then the KK spectrum of these curves will not form complete GUT multiplets. The heavy modes exchanged will not be full GUT representations but rather there will be a mass splitting between the doublets exchanged and the triplets exchanged. This means that there will be a different suppression for the quark Yukawas compared to the lepton Yukawas. The reason for the mass splitting (\ref{gjmass}) then follows straightforwardly. We are unable to explain exactly the factors of 3 but this mechanism does explain why the heaviest generation does not have a mass splitting while the lighter ones do: it is the light generations masses that are sensitive to KK-scale non-GUT physics induced by the hypercharge flux.\footnote{It may be possible to explain why the mass ratios are inverse for the two lighter generations due to opposite restrictions of the hypercharge flux so that on one curve a doublet is the lightest KK mode while on the other a triplet is the lightest KK mode. Note however that this is not the case for the model in table \ref{tab:model21113cur}.}

\section{Summary}
\label{sec:summary}

We have studied $SU(5)$ F-theory GUT models in a semi-local framework that are based on small monodromy groups with multiple $U(1)$ factors. We constructed explicit semi-local geometries and determined the homology classes of the matter curves. We then studied phenomenological aspects of these constructions. In particular we studied the effects of the hypercharge flux on the matter spectrum.

Perhaps the most important conclusion from this work is that in these models, after the exotics are lifted, the $U(1)$ symmetries are not sufficient by themselves to completely forbid dangerous proton decay operators. Indeed we found that some extra selection principle is needed which we took to be a version of R-parity. It may be that some other selection principle can be utilised such as separating curves and using geometric wavefunction suppression. Either way this is an important component of such model building and deserves further study.

For models where all the generations were localised on a single curve we have given explicit models, for all the monodromy groups, where all the exotics induced are lifted to a high mass scale through singlet vevs while proton decay operators are not generated. We have shown that gauge coupling unification can be maintained to the same accuracy as the MSSM even with the exotics below the GUT scale due to the particular spectrum induced by the hypercharge flux. We also presented viable neutrino scenarios for these models. The case where the generations come from different curves was also successful in these aspects but gauge coupling unification was not quite as accurate though still within the size of the required threshold corrections of the MSSM. Finally, for these latter models, we presented a mechanism that combines the Froggatt-Nielsen approach to flavour with hypercharge flux GUT breaking and can account for quark-lepton mass splitting at the GUT scale.

We have not commented so far on supersymmetry breaking. There is a non-trivial problem with the size of the $\mu$-term since in most of our models the $\mu$-term as generated by the Giudice-Masiero mechanism was suppressed relative to the gravitino mass by factors of the singlet vevs. Such a suppression is phenomenologically unfavoured for both gravity and gauge mediation. If the singlet vevs are taken large enough this problem can be eased. Notice also that exotics discussed in our paper are natural candidates for messenger fields
in gauge mediation of supersymmetry breaking. 

The most obvious and important continuation of this work is to attempt to construct global realisations. It should be expected that such global realisations, if possible to find, would be much more constrained than the semi-local models studied here. In particular a global realisation is essential for determining the $U(1)$ fluxes that lead to the chirality and also for finding any possible geometric symmetries that can play the crucial role of the R-parity we have imposed by hand.

\subsection*{Acknowledgments}

We thank Jonathan Heckman, Joe Marsano, Graham Ross, Sakura Schafer-Nameki, Cumrun Vafa and Timo Weigand for stimulating and helpful discussions and explanations.

The work presented was supported in part by the European ERC Advanced Grant 226371 MassTeV, by the CNRS PICS no. 3059 and 4172,
by the grants ANR-05-BLAN-0079-02, the PITN contract PITN-GA-2009-237920 and the IFCPAR CEFIPRA programme 4104-2.

\appendix

\section{More single curve models}
\label{sec:moersingcur}

In this appendix we present models where doublet-triplet splitting is performed directly by the hypercharge flux. These tend to have more exotics than the models presented in the main text. However since there are no exotics on the Higgs curves we can realise the R-parity assignments of all the singlets having positive R-parity and all the matter apart from the Higgs curves having negative parity. As discussed in section \ref{sec:rparity}, this assignment always forbids proton decay if the stronger version of R-parity is imposed.

\subsection{Models based on $2+1+1+1$ monodromy}
\label{sec:moersingcur1}

A model that realises direct splitting by flux is shown in table \ref{tab:model2111app}. We have taken the hypercharge flux restrictions $N_7=-1$, $N_8=2$ and $N_9=0$.
\begin{table}[ht]
\center
{\scriptsize
\begin{tabular}{|cccccc|}
\hline
Field & Curve & $N_{Y}$ & $M_{U(1)}$ & Exotics & R-parity  \\
\hline
$\f_{H_u}$   & $-2t_1$ &  +1 & 0 &  & + \\
$\fb_{H_d}$  & $t_3+t_5$  &  -1 & 0  &  & +  \\
$\fb_{M}$    & $t_1+t_4$  &  -1 & -3 & $(1,2)_{-1/2}$ & - \\
$\fb_{1}$    & $t_1+t_3$ &  -1 & -1 & $(\bar{3},1)_{+1/3}+2\times(1,2)_{-1/2}$ & - \\
$\fb_{3}$    & $t_1+t_5$  &  -1 & 0 & $(1,2)_{-1/2}$ & -  \\
$\f_{4}$     & $-t_3-t_4$ &  +1 & +1 & $(3,1)_{-1/3}+2\times(1,2)_{+1/2}$ & - \\
$\f_{6}$     & $-t_4-t_5$  &  +2 & 0 & $2\times(1,2)_{+1/2}$ & -  \\
$\te_{M}$    & $t_1$      &  -1 & +4 & $(3,2)_{+1/6} + 2\times(\bar{3},1)_{-2/3}$ & - \\
$\te_{2}$    & $t_3$     &  -1 & +1 & $(3,2)_{+1/6} + 2\times(\bar{3},1)_{-2/3}$ & - \\
$\teb_{3}$   & $-t_4$      &  +2 & -2 & $2\times(\bar{3},2)_{-1/6} + 4\times(3,1)_{+2/3}$ & - \\
$\te_{4}$    & $t_5$      &  0 & 0 &  &  \\
\hline \hline
Singlet & Curve & vev & F-term & & R-parity  \\
\hline
$X_1$ & $-t_1+t_4$ & $\epsilon_1$ &  & & + \\
$X_2$ & $t_1-t_3$ & $\epsilon_2$ &  & & +  \\
$X_3$ & $-t_1+t_5$ & $\epsilon_3$ & $\left<F_3\right>$ & & +  \\
\hline \hline
Induced mass & \multicolumn{4}{c}{Exotics lifted} & R-parity \\
\hline
$\epsilon_3 \fb_M \f_6$ & \multicolumn{4}{c}{$(1,2)_{-1/2}(1,2)_{+1/2}$} & + \\
$\epsilon_1 \fb_1 \f_4$ & \multicolumn{4}{c}{$(3,1)_{-1/3}(\bar{3},1)_{+1/3}+2\times(1,2)_{-1/2}(1,2)_{+1/2}$} & + \\
$\epsilon_1 \fb_3 \f_6$ & \multicolumn{4}{c}{$(1,2)_{-1/2}(1,2)_{+1/2}$} & + \\
$\epsilon_1 \te_M \teb_3$ & \multicolumn{4}{c}{$(3,2)_{+1/6}(\bar{3},2)_{-1/6}+2\times(\bar{3},1)_{-2/3}(3,1)_{+2/3}$} & + \\
$\epsilon_1\epsilon_2 \te_2 \teb_3$ & \multicolumn{4}{c}{$(3,2)_{+1/6}(\bar{3},2)_{-1/6}+2\times(\bar{3},1)_{-2/3}(3,1)_{+2/3}$} & + \\
\hline \hline
Operator & Charges & Super/Kahler potential & \multicolumn{2}{c}{$U(1)$ Neutrality} & R-parity\\
\hline
$\f_{H_u}\fb_{H_d}$ & $-2t_1 + t_3+t_5$ & W & \multicolumn{2}{c}{} & +\\
$\fb_{M}\fb_{M}\te_M$ & $3t_1+2t_4$ & W & \multicolumn{2}{c}{} & - \\
$\te_{M}\te_{M}\te_{M}\fb_{M}$ & $4t_1+t_4$ & W & \multicolumn{2}{c}{} & + \\
$\f_{H_u}\fb_{M}$ & $-t_1+t_4$ & W & \multicolumn{2}{c}{} & - \\
$\left<F^{\dagger}\right>\f_{H_u}\fb_{H_d}$ & $-2t_1 +t_3+t_5$ & K & \multicolumn{2}{c}{$\epsilon_2\left<F^{\dagger}_3\right>\f_{H_u}\fb_{H_d}$} & + \\
\hline
\end{tabular}
}
\caption{Table showing flux restrictions, induced exotics, singlet vevs and induced operators with positive powers of singlet insertions for a model based on a $2+1+1+1$ splitting where doublet-triplet splitting is done directly by flux.}
\label{tab:model2111app}
\end{table}

Consider first gauge coupling unification. We find that putting the exotics at a scale $2\times10^{15}$GEV the gauge couplings at the GUT scale read $\alpha^{-1}_1=20.8$, $\alpha^{-1}_2=20.7$ and $\alpha^{-1}_3=20.9$. This is nearly as good as 1-loop MSSM unification accuracy. The reason for this is that the exotics of the $\te$ and the $\f$ representations, although not complete GUT representations in themselves, when added act as 2 complete vector pairs of $\te$ representations, 4 complete vector pairs of $\f$ representations and one vector pair of doublets as far as the gauge coupling running is concerned (\ref{specexot1}).

We see from table \ref{tab:model2111app} that a $\mu$-term is forbidden by the $U(1)$ charges, and that a Giudice-Masiero mass can be induced by an appropriate F-term. Also the problematic $\f_{H_u}\fb_M$ coupling is forbidden by the $U(1)$ charges.

We now turn to proton decay. First we note that dimension 4 and 5 proton decay operators involving positive powers of singlet vevs are forbidden due to the $U(1)$ symmetries. However they could still be potentially induced by integrating out states which is why R-parity is required as discussed in section \ref{sec:rparity}. It is instructive to see how it acts in this particular case and how otherwise proton decay would be induced.

In the absence of R-parity there is danger of the proton decay operator being generated by integrating out charged triplets. The Higgs triplets can not mediate proton decay since they require a $\mu$-term coupling which is forbidden. Also the massless exotic triplets on the $\fb_1$ and $\f_4$ curves that gain a mass from the singlets can not mediate proton decay because the coupling $\f_4\te_M\te_M$ is forbidden by the $U(1)$ symmetries. However triplet KK modes along other exotic curves can mediate proton decay. An example of how such a mediation can occur is by the operators
\be
W \supset \fb_{H_d}^{KK} \fb_M \te_M + \epsilon_2 \fb_1^{KK} \f^{KK}_{H_u} + \f^{KK}_{H_u} \te_M \te_M + \epsilon_3 \fb_1^{KK} \f_{H_d}^{KK}\;. \label{tripkkproton}
\ee
Integrating out the KK modes gives a proton decay operator
\be
\left(\frac{\epsilon_3}{\epsilon_2}\right) \fb_M \te_M \te_M \te_M \;. \label{2111prodecop}
\ee
In this case the R-parity assignment as in table \ref{tab:model2111app} forbids the 2nd and 4th operators of (\ref{tripkkproton}). This is an example case of the more general result that such an R-parity forbids proton decay presented in section \ref{sec:rparity}.\footnote{It is also possible to forbid proton decay using the weaker version of R-parity by assigning $\epsilon_2$ negative charge and $\te_2$ positive charge. However then a $\mu$-term is consequently forbidden and it is not possible to induce it by introducing an extra singlet without also inducing proton decay.}

Finally we turn to the neutrino scenarios. Both can be realised by taking the right handed neutrino curve to be $N=-t_4+t_5$ for Dirac and $N=t_1-t_4$ for Majorana. However there is the usual compatibility problem between the Dirac scenario and the Giudice-Masiero mechanism. For example a Dirac scenario would involve a right-handed neutrino choice of $N=-t_4+t_5$. This can not appear in the superpotential and so there are no superpotential Dirac or Majorana masses which allows for the Kahler potential Dirac scenario. Indeed such Dirac mass terms are obtained through the Kahler potential operator $\epsilon_2\f_{H_d}\fb_{M}N$. However we also have the Kahler potential operator
\be
K \supset \epsilon_1^2\epsilon_3\left<F^{\dagger}_3\right> N N \;,
\ee
which leads to TeV scale superpotential Majorana masses that suppress the neutrino masses too much. Unless this operator is somehow suppressed (perhaps geometrically) or the F-term is vanishing, in which case we give up on the Giudice-Masiero mechanism, this scenario is not viable.\footnote{Note that it is possible to generate the $\mu$-term through a singlet vev, for example $X=t_3-t_5$, directly at the superpotential level as in the NMSSM. Since the singlet has only a TeV vev it does not induce large enough dimension 5 proton decay to be a problem.}

We can also realise a Majorana scenario with charged GUT singlets. For example taking the right-handed neutrino curve to be $N=t_1-t_4$. The resulting neutrino masses are given by the operators
\be
W \supset \f_{H_u} \fb_M N + \epsilon_1^2 NN \;,
\ee
which give the mass scale
\be
M_{\nu} \sim  \frac{1}{\epsilon_1^2} 10^{-3} \mathrm{eV} \;,
\ee
which is phenomenologically viable as long as $\epsilon_1$ is not too small. Note that, as discussed in section \ref{sec:singcur}, there is a linear term
\be
W \supset \epsilon_1 N \;,
\ee
which must be forbidden by R-parity by assigning negative charge to the right handed neutrino.

\subsection*{Model with different matter curves}

Here we present a model which allows for the Giudice-Masiero term $X^{\dagger}\f_{H_u}\fb_{H_d}$ which gives a preferably heavier $\mu$-term. However this model also has dimension 5 proton decay operators induced which should be forbidden by R-parity.
The model is based on a $2+1+1+1$ monodromy group and is shown in table \ref{tab:modelleastexot21113}. It has hypercharge flux choices $N_7=+1$, $N_8=0$, $N_9=0$.
\begin{table}[ht]
\center
{\scriptsize
\begin{tabular}{|cccccc|}
\hline
Field & Curve & $N_{Y}$ & $M_{U(1)}$ & Exotics & R-parity \\
\hline
$\f_{H_u}$   & $-2t_1$    &  +1  & 0  &  & + \\
$\fb_{H_d}$  & $t_1+t_3$  &  -1 & 0 &  & + \\
$\fb_{M}$    & $t_4+t_5$  &  0  & -3 &  & - \\
$\fb_{2}$    & $t_1+t_4$  &  -1  & 0 & $(1,2)_{-1/2}$ & - \\
$\fb_{3}$    & $t_1+t_5$  &  -1  & -4 & $4\times(\bar{3},1)_{+1/3} + 5\times(1,2)_{-1/2}$ & - \\
$\f_{4}$     & $-t_3-t_4$ &  +1  & +4 & $4\times(3,1)_{-1/3} + 5\times(1,2)_{+1/2}$ & - \\
$\f_{5}$    & $-t_3-t_5$   &  +1   & 0 & $(1,2)_{+1/2}$ & - \\
$\te_{M}$    & $t_1$      &  -1  & +4 & $(3,2)_{+1/6} + 2\times(\bar{3},1)_{-2/3}$ & - \\
$\teb_{2}$    & $-t_3$     &  +1 & -1 & $(\bar{3},2)_{-1/6} + 2\times(3,1)_{+2/3}$ & - \\
$\te_{3}$    & $t_4$      &  0 & 0 &  &  \\
$\te_{4}$    & $t_5$      &  0 & 0 &  &  \\
\hline \hline
Singlet & Curve & vev & F-term & & R-parity \\
\hline
$X_1$ & $-t_1+t_3$ & $\epsilon_1$ & $\left<F_1\right>$ &  & + \\
$X_2$ & $t_4-t_5$  & $\epsilon_2$ &  &  & + \\
\hline \hline
Induced mass & \multicolumn{4}{c}{Exotics lifted} & R-parity\\
\hline
$\epsilon_1 \f_{4} \fb_2$ & \multicolumn{4}{c}{$(1,2)_{-1/2}(1,2)_{+1/2}$} & + \\
$\epsilon_1\epsilon_2 \f_{4} \fb_3$ & \multicolumn{4}{c}{$4\times(3,1)_{-1/3}(\bar{3},1)_{+1/3}+4\times(1,2)_{+1/2}(1,2)_{-1/2}$} & + \\
$\epsilon_1 \f_{5} \fb_3$ & \multicolumn{4}{c}{$(1,2)_{-1/2}(1,2)_{+1/2}$} & + \\
$\epsilon_1 \te_M \teb_2$ & \multicolumn{4}{c}{$(3,2)_{+1/6}(\bar{3},2)_{-1/6} + 2\times(\bar{3},1)_{-2/3}(3,1)_{+2/3}$} & + \\
\hline \hline
Operator & Charges & Super/Kahler potential & \multicolumn{2}{c}{$U(1)$ Neutrality} & R-parity\\
\hline
$\f_{H_u}\fb_{H_d}$ & $-t_1 +t_3$ & W & \multicolumn{2}{c}{} & + \\
$\fb_{M}\fb_{M}\te_M$ & $3t_1+2t_4$ & W & \multicolumn{2}{c}{} & - \\
$\te_{M}\te_{M}\te_{M}\fb_{M}$ & $4t_1+t_4$ & W & \multicolumn{2}{c}{$\epsilon_1\te_{M}\te_{M}\te_{M}\fb_{M}$} & + \\
$\f_{H_u}\fb_{M}$ & $-t_1+t_4$ & W & \multicolumn{2}{c}{} & - \\
$\left<F^{\dagger}\right>\f_{H_u}\fb_{H_d}$ & $-2t_1 +t_3+t_5$ & K & \multicolumn{2}{c}{$\left<F^{\dagger}_1\right>\f_{H_u}\fb_{H_d}$} & + \\
\hline
\end{tabular}
}
\caption{Table showing flux restrictions, induced exotics, singlet vevs and induced operators with positive powers of singlet insertions for a model based on a $2+1+1+1$ splitting.}
\label{tab:modelleastexot21113}
\end{table}

The exotics act as 5 vector pairs of $\f$ representations 1 vector pair of $\te$ representations and one doublet pair (\ref{specexot1}). We find that putting the exotics at a scale $2\times10^{15}$GEV the gauge couplings at the GUT scale read $\alpha^{-1}_1=21.2$, $\alpha^{-1}_2=21.0$ and $\alpha^{-1}_3=21.3$.

\subsection{Models based on $2+2+1$ monodromy}
\label{sec:moersingcur2}

Since we require doublet-triplet splitting to be done directly by flux rather than by the singlet vevs, the hypercharge restrictions are fixed to be $N_1=-1$ and $N_2=+3$. The minimal such model is shown in table \ref{tab:model221main}.
\begin{table}[ht]
\center
{\scriptsize
\begin{tabular}{|cccccc|}
\hline
Field & Curve & $N_{Y}$ & $M_{U(1)}$ & Exotics & R-parity \\
\hline
$\f_{H_u}$   & $-2t_1$    &  +1   & 0 &  & + \\
$\fb_{H_d}$  & $t_3+t_5$  &  -1  & 0  &  & + \\
$\fb_{M}$    & $t_1+t_3$  &  -2   & -3 & $2\times(1,2)_{-1/2}$ & - \\
$\fb_{2}$     & $t_1+t_5$ &  -1   & -1  & $(\bar{3},1)_{+1/3}+2\times(1,2)_{-1/2}$ & - \\
$\f_{4}$    & $-2t_3$     &  +3  & +1 & $(3,1)_{-1/3}+4\times(1,2)_{+1/2}$ & - \\
$\te_{M}$    & $t_1$      &  -1   & +4 & $(3,2)_{+1/6} + 2\times(\bar{3},1)_{-2/3}$ & - \\
$\teb_{2}$    & $-t_3$     &  +3  & -3 & $3\times(\bar{3},2)_{-1/6} + 6\times(3,1)_{+2/3}$ & - \\
$\te_{3}$    & $t_5$    &  -2  & +2 & $2\times(3,2)_{+1/6} + 4\times(\bar{3},1)_{-2/3}$ & - \\
\hline \hline
Singlet & Curve & vev & F-term & & R-parity \\
\hline
$X_1$ & $-t_1+t_3$ & $\epsilon_1$ & $\left<F_1\right>$ & & +  \\
$X_2$ & $t_3-t_5$ & $\epsilon_2$ &  & & +  \\
\hline \hline
Induced mass & \multicolumn{4}{c}{Exotics lifted} & R-parity\\
\hline
$\epsilon_1 \fb_{M} \f_4$ & \multicolumn{4}{c}{$2\times(1,2)_{+1/2}(1,2)_{-1/2}$} & + \\
$\epsilon_1\epsilon_2 \fb_{2} \f_4$ & \multicolumn{4}{c}{$(3,1)_{-1/3}(\bar{3},1)_{+1/3} + 2\times(1,2)_{+1/2}(1,2)_{-1/2}$} & + \\
$\epsilon_1 \te_M \teb_2$ & \multicolumn{4}{c}{$(\bar{3},2)_{-1/6}(3,2)_{+1/6} + 2\times(3,1)_{+2/3}(\bar{3},1)_{-2/3}$} & + \\
$\epsilon_2 \te_3 \teb_2$ & \multicolumn{4}{c}{$2\times(\bar{3},2)_{-1/6}(3,2)_{+1/6} + 4\times(3,1)_{+2/3}(\bar{3},1)_{-2/3}$} & + \\
\hline \hline
Operator & Charges & Super/Kahler potential & \multicolumn{2}{c}{$U(1)$ Neutrality} & R-parity \\
\hline
$\f_{H_u}\fb_{H_d}$ & $-2t_1 +t_3+t_5$ & W & \multicolumn{2}{c}{} & + \\
$\fb_{M}\fb_{M}\te_M$ & $3t_1+2t_3$ & W & \multicolumn{2}{c}{} & - \\
$\te_{M}\te_{M}\te_{M}\fb_{M}$ & $4t_1+t_3$ & W & \multicolumn{2}{c}{} & + \\
$\f_{H_u}\fb_{M}$ & $-t_1+t_3$ & W & \multicolumn{2}{c}{} & - \\
$\left<F^{\dagger}\right>\f_{H_u}\fb_{H_d}$ & $-2t_1 +t_3+t_5$ & K & \multicolumn{2}{c}{$\epsilon_1\epsilon_2\left<F^{\dagger}_1\right>\f_{H_u}\fb_{H_d}$} & + \\
\hline
\end{tabular}
}
\caption{Table showing flux restrictions, induced exotics, singlet vevs and induced operators for a model based on a $2+2+1$ splitting.}
\label{tab:model221main}
\end{table}

Consider first gauge coupling unification. We find that putting the exotics at a scale $2\times10^{15}$GEV the gauge couplings at the GUT scale read $\alpha^{-1}_1=19.5$, $\alpha^{-1}_2=19.5$ and $\alpha^{-1}_3=19.4$. This is equivalent to 1-loop MSSM unification accuracy. The reason for this is that the exotics act as 3 complete vector pairs of $\te$ and 4 complete vector pairs of $\f$ representations as far as the gauge coupling running is concerned
\bea
 \left[\te + \f\right] &=& \left[ (3,2)_{+1/6} + (\bar{3},1)_{-2/3} + (1,1)_{+1} + (3,1)_{-1/3} + (1,2)_{+1/2}\right] \nn \\
 &\sim & \left[ (3,2)_{+1/6} + (\bar{3},1)_{-2/3} + (3,1)_{+2/3} + (1,2)_{+1/2}\right] \;. \label{specexot1}
\eea
We find that such combinations arise frequently and naturally in these models.

From table \ref{tab:model221main} we see that a $\mu$-term is forbidden by the $U(1)$ charges and a Giudice-Masiero term can be induced by the appropriate F-term. The dangerous term $\f_{H_u}\fb_{M}$ is forbidden by the $U(1)$ charges. However as in section \ref{sec:models2111} proton decay operators can be induced by integrating out KK modes. This problem implies that R-parity must be imposed by hand as presented in table \ref{tab:model221main}.

We now turn to the neutrinos. It is possible to realise the Dirac scenario with the right handed neutrino being $N=t_1-t_5$. In this case a Dirac mass is induced through the Kahler potential operator $(\epsilon_1\epsilon_2)^2\f_{H_d}\fb_M N$. There is no superpotential Dirac or Majorana mass because of the $-t_5$ factor. However, as usual, if we induce a Giudice-Masiero mass the singlet F-term also induces a TeV scale Majorana mass through the Kahler potential operator $\epsilon_1 (\epsilon_1\epsilon_2)^2 \left<F_1^{\dagger}\right> NN$. Therefore the two mechanisms are incompatible.

It is possible to realise a Majorana scenario by taking $N=t_1-t_3$ which gives neutrino masses of
\be
M_{\nu} \sim  \frac{1}{\epsilon_1^2} 10^{-3} \mathrm{eV} \;.
\ee

\subsection*{Model with different matter curves}

In this final section we discuss a model based on $2+2+1$ monodromy which has different choices for the matter curves than those determined in section \ref{sec:models221}. As noted such a choice leads to proton decay operators being allowed by the $U(1)$ symmetries. However if we impose the stronger version of R-parity such operators are forbidden.

The model is shown in table \ref{tab:model221main2} and has $N_1=-1$ and $N_2=0$.
\begin{table}[ht]
\center
{\scriptsize
\begin{tabular}{|cccccc|}
\hline
Field & Curve & $N_{Y}$ & $M_{U(1)}$ & Exotics & R-parity \\
\hline
$\f_{H_u}$   & $-2t_1$    &  +1   & 0  &  & + \\
$\fb_{H_d}$  & $t_1+t_5$  &  -1   & 0  &  & + \\
$\fb_{M}$    & $2t_3$     &  0   & -3 &  & - \\
$\fb_{1}$     & $t_1+t_3$ &  -2   & -4 & $4\times(\bar{3},1)_{+1/3}+6\times(1,2)_{-1/2}$ & - \\
$\f_{3}$    & $-t_3-t_5$  &  +2   & +4 & $4\times(3,1)_{-1/3}+6\times(1,2)_{+1/2}$ & - \\
$\te_{M}$    & $t_1$      &  -1   & +4 & $(3,2)_{+1/6} + 2\times(\bar{3},1)_{-2/3}$ & - \\
$\te_{2}$    & $t_3$      &  0   & 0 &  &  \\
$\teb_{3}$    & $-t_5$    &  +1   & -1 & $(\bar{3},2)_{-1/6} + 2\times(3,1)_{+2/3}$ & - \\
\hline \hline
Singlet & Curve & vev & F-term & & R-parity \\
\hline
$X_1$ & $-t_1+t_5$ & $\epsilon_1$ &  & & +  \\
\hline \hline
Induced mass & \multicolumn{4}{c}{Exotics lifted} & R-parity\\
\hline
$\epsilon_1 \fb_{1} \f_3$ & \multicolumn{4}{c}{$4\times(\bar{3},1)_{+1/3}(3,1)_{-1/3}+6\times(1,2)_{-1/2}(1,2)_{+1/2}$} & + \\
$\epsilon_1 \te_M \teb_2$ & \multicolumn{4}{c}{$(\bar{3},2)_{-1/6}(3,2)_{+1/6} + 2\times(3,1)_{+2/3}(\bar{3},1)_{-2/3}$} & + \\
\hline \hline
Operator & Charges & Super/Kahler potential & \multicolumn{2}{c}{$U(1)$ Neutrality} & R-parity \\
\hline
$\f_{H_u}\fb_{H_d}$ & $-t_1+t_5$ & W & \multicolumn{2}{c}{} & + \\
$\fb_{M}\fb_{M}\te_M$ & $t_1+4t_3$ & W & \multicolumn{2}{c}{} & - \\
$\te_{M}\te_{M}\te_{M}\fb_{M}$ & $3t_1+2t_3$ & W & \multicolumn{2}{c}{$\epsilon_1\te_{M}\te_{M}\te_{M}\fb_{M}$} & + \\
$\f_{H_u}\fb_{M}$ & $-2t_1+2t_3$ & W & \multicolumn{2}{c}{} & - \\
$\left<F^{\dagger}\right>\f_{H_u}\fb_{H_d}$ & $-t_1+t_5$ & K & \multicolumn{2}{c}{$\left<F^{\dagger}_1\right>\f_{H_u}\fb_{H_d}$} & + \\
\hline
\end{tabular}
}
\caption{Table showing flux restrictions, induced exotics, singlet vevs and induced operators for a model based on a $2+2+1$ splitting.}
\label{tab:model221main2}
\end{table}
The exotics act as 5 vector pairs of $\f$ representations, 1 vector pair of $\te$ representations and 1 pair of doublets. This means that placing them at $2\times10^{15}$GeV gives $\alpha^{-1}_1=21.2$, $\alpha^{-1}_2=21.0$ and $\alpha^{-1}_3=21.3$.



\begin{thebibliography}{99}

\bibitem{Donagi:2008ca}
  R.~Donagi and M.~Wijnholt,
  ``Model Building with F-Theory,''
  arXiv:0802.2969 [hep-th].

\bibitem{Beasley:2008dc}
  C.~Beasley, J.~J.~Heckman and C.~Vafa,
  ``GUTs and Exceptional Branes in F-theory - I,''
  JHEP {\bf 0901} (2009) 058
  [arXiv:0802.3391 [hep-th]].

\bibitem{Beasley:2008kw}
  C.~Beasley, J.~J.~Heckman and C.~Vafa,
  ``GUTs and Exceptional Branes in F-theory - II: Experimental Predictions,''
  JHEP {\bf 0901} (2009) 059
  [arXiv:0806.0102 [hep-th]].

\bibitem{Donagi:2008kj}
  R.~Donagi and M.~Wijnholt,
  ``Breaking GUT Groups in F-Theory,''
  arXiv:0808.2223 [hep-th].

\bibitem{Heckman:2010bq}
  J.~J.~Heckman,
  ``Particle Physics Implications of F-theory,''
  arXiv:1001.0577 [hep-th].

\bibitem{Buican:2006sn}
  M.~Buican, D.~Malyshev, D.~R.~Morrison, H.~Verlinde and M.~Wijnholt,
  ``D-branes at singularities, compactification, and hypercharge,''
  JHEP {\bf 0701} (2007) 107
  [arXiv:hep-th/0610007].

\bibitem{Blumenhagen:2008aw}
  R.~Blumenhagen,
  ``Gauge Coupling Unification In F-Theory Grand Unified Theories,''
  Phys.\ Rev.\ Lett.\  {\bf 102} (2009) 071601
  [arXiv:0812.0248 [hep-th]].

\bibitem{Marsano:2009gv}
J.~Marsano, N.~Saulina and S.~Schafer-Nameki,
``Monodromies, Fluxes, and Compact Three-Generation F-theory GUTs,''
arXiv:0906.4672 [hep-th].

\bibitem{Bouchard:2009bu}
  V.~Bouchard, J.~J.~Heckman, J.~Seo and C.~Vafa,
  ``F-theory and Neutrinos: Kaluza-Klein Dilution of Flavor Hierarchy,''
  arXiv:0904.1419 [hep-ph].

\bibitem{Heckman:2009mn}
  J.~J.~Heckman, A.~Tavanfar and C.~Vafa,
  ``The Point of E8 in F-theory GUTs,''
  arXiv:0906.0581 [hep-th].

\bibitem{Marsano:2009wr}
  J.~Marsano, N.~Saulina and S.~Schafer-Nameki,
  ``Compact F-theory GUTs with $U(1)_{PQ}$,''
  arXiv:0912.0272 [hep-th].

\bibitem{Dudas:2009hu}
  E.~Dudas and E.~Palti,
  ``Froggatt-Nielsen models from E8 in F-theory GUTs,''
  JHEP {\bf 1001}, 127 (2010)
  [arXiv:0912.0853 [hep-th]].

\bibitem{King:2010mq}
  S.~F.~King, G.~K.~Leontaris and G.~G.~Ross,
  ``Family symmetries in F-theory GUTs,''
  arXiv:1005.1025 [hep-ph].

\bibitem{Heckman:2010xz}
  J.~J.~Heckman, J.~Shao and C.~Vafa,
  ``F-theory and the LHC: Stau Search,''
  arXiv:1001.4084 [hep-ph].

\bibitem{Hayashi:2009ge}
  H.~Hayashi, T.~Kawano, R.~Tatar and T.~,
  ``Codimension-3 Singularities and Yukawa Couplings in F-theory,''
  Nucl.\ Phys.\  B {\bf 823} (2009) 47
  [arXiv:0901.4941 [hep-th]].

\bibitem{Donagi:2009ra}
  R.~Donagi and M.~Wijnholt,
  ``Higgs Bundles and UV Completion in F-Theory,''
  arXiv:0904.1218 [hep-th].

\bibitem{Hayashi:2010zp}
  H.~Hayashi, T.~Kawano, Y.~Tsuchiya and T.~Watari,
  ``More on Dimension-4 Proton Decay Problem in F-theory -- Spectral Surface,
  Discriminant Locus and Monodromy,''
  arXiv:1004.3870 [hep-th].

\bibitem{Chen:2010tp}
  C.~M.~Chen and Y.~C.~Chung,
  ``Flipped SU(5) GUTs from E8 Singularity in F-theory,''
  arXiv:1005.5728 [hep-th].

\bibitem{Giudice:1988yz}
  G.~F.~Giudice and A.~Masiero,
  ``A Natural Solution to the mu Problem in Supergravity Theories,''
  Phys.\ Lett.\  B {\bf 206} (1988) 480.

\bibitem{Georgi:1979df}
  H.~Georgi and C.~Jarlskog,
  ``A New Lepton - Quark Mass Relation In A Unified Theory,''
  Phys.\ Lett.\  B {\bf 86} (1979) 297.

\bibitem{Ross:2007az}
  G.~Ross and M.~Serna,
  ``Unification and Fermion Mass Structure,''
  Phys.\ Lett.\  B {\bf 664} (2008) 97
  [arXiv:0704.1248 [hep-ph]].

\bibitem{Blumenhagen:2008zz}
  R.~Blumenhagen, V.~Braun, T.~W.~Grimm and T.~Weigand,
  ``GUTs in Type IIB Orientifold Compactifications,''
  Nucl.\ Phys.\  B {\bf 815} (2009) 1
  [arXiv:0811.2936 [hep-th]].

\bibitem{09024143}
  B.~Andreas and G.~Curio,
  ``From Local to Global in F-Theory Model Building,''
  arXiv:0902.4143 [hep-th].

\bibitem{09060013}
  R.~Blumenhagen, T.~W.~Grimm, B.~Jurke and T.~Weigand,
  ``F-theory uplifts and GUTs,''
  JHEP {\bf 0909} (2009) 053
  [arXiv:0906.0013 [hep-th]].

\bibitem{Blumenhagen:2009yv}
  R.~Blumenhagen, T.~W.~Grimm, B.~Jurke and T.~Weigand,
  ``Global F-theory GUTs,''
  arXiv:0908.1784 [hep-th].

\bibitem{09043932}
  J.~Marsano, N.~Saulina and S.~Schafer-Nameki,
  ``F-theory Compactifications for Supersymmetric GUTs,''
  JHEP {\bf 0908} (2009) 030
  [arXiv:0904.3932 [hep-th]].

\bibitem{Grimm:2009yu}
  T.~W.~Grimm, S.~Krause and T.~Weigand,
  ``F-Theory GUT Vacua on Compact Calabi-Yau Fourfolds,''
  arXiv:0912.3524 [hep-th].

\bibitem{Grimm:2010ez}
  T.~W.~Grimm and T.~Weigand,
  ``On Abelian Gauge Symmetries and Proton Decay in Global F-theory GUTs,''
  arXiv:1006.0226 [hep-th].

\bibitem{Marsano:2010ix}
  J.~Marsano, N.~Saulina and S.~Schafer-Nameki,
  ``A Note on G-Fluxes for F-theory Model Building,''
  arXiv:1006.0483 [hep-th].

\bibitem{Hayashi:2009bt}
  H.~Hayashi, T.~Kawano, Y.~Tsuchiya and T.~Watari,
  ``Flavor Structure in F-theory Compactifications,''
  arXiv:0910.2762 [hep-th].

\bibitem{morrisonkatz}
  S.~Katz,, D.~R.~Morrison,
  ``Gorenstein Threefold Singularities with Small Resolutions via Invariant Theory for Weyl Groups,''
  J. Alg. Geom. 1 (1992) 449--530
  [arXiv:alg-geom/9202002v1].

\bibitem{Bershadsky:1996nh}
  M.~Bershadsky, K.~A.~Intriligator, S.~Kachru, D.~R.~Morrison, V.~Sadov and C.~Vafa,
  ``Geometric singularities and enhanced gauge symmetries,''
  Nucl.\ Phys.\  B {\bf 481} (1996) 215
  [arXiv:hep-th/9605200].

\bibitem{Katz:1996xe}
  S.~H.~Katz and C.~Vafa,
  ``Matter from geometry,''
  Nucl.\ Phys.\  B {\bf 497}, 146 (1997)
  [arXiv:hep-th/9606086].

\bibitem{Heckman:2008qa}
  J.~J.~Heckman and C.~Vafa,
  ``Flavor Hierarchy From F-theory,''
  arXiv:0811.2417 [hep-th].

\bibitem{Font:2009gq}
  A.~Font and L.~E.~Ibanez,
  ``Matter wave functions and Yukawa couplings in F-theory Grand Unification,''
  arXiv:0907.4895 [hep-th].

\bibitem{Cecotti:2009zf}
  S.~Cecotti, M.~C.~N.~Cheng, J.~J.~Heckman and C.~Vafa,
  ``Yukawa Couplings in F-theory and Non-Commutative Geometry,''
  arXiv:0910.0477 [hep-th].

\bibitem{Conlon:2009qq}
  J.~P.~Conlon and E.~Palti,
  ``Aspects of Flavour and Supersymmetry in F-theory GUTs,''
  arXiv:0910.2413 [hep-th].


\bibitem{Marchesano:2009rz}
  F.~Marchesano and L.~Martucci,
  ``Non-perturbative effects on seven-brane Yukawa couplings,''
  arXiv:0910.5496 [hep-th].

\bibitem{Smirnov:1996bg}
  A.~Y.~Smirnov and F.~Vissani,
  ``Upper bound on all products of R-parity violating couplings $\lambda'$ and
  $\lambda''$ from proton decay,''
  Phys.\ Lett.\  B {\bf 380} (1996) 317
  [arXiv:hep-ph/9601387].

\bibitem{salatil}
  P.~Salati1 and J.~C.~Wallet2
  ``Proton and neutron decay rates in conventional and supersymmetric guts,''
  Nucl.\ Phys.\  B {\bf 209} (1982) 389.

\bibitem{Ellis:1983qm}
  J.~R.~Ellis, J.~S.~Hagelin, D.~V.~Nanopoulos and K.~Tamvakis,
  ``Observable Gravitationally Induced Baryon Decay,''
  Phys.\ Lett.\  B {\bf 124} (1983) 484.

\bibitem{Ibanez:1991pr}
  L.~E.~Ibanez and G.~G.~Ross,
  ``Discrete Gauge Symmetries And The Origin Of Baryon And Lepton Number
  Conservation In Supersymmetric Versions Of The Standard Model,''
  Nucl.\ Phys.\  B {\bf 368} (1992) 3.

\bibitem{Barbier:2004ez}
  R.~Barbier {\it et al.},
  ``R-parity violating supersymmetry,''
  Phys.\ Rept.\  {\bf 420} (2005) 1
  [arXiv:hep-ph/0406039].

\bibitem{Conlon:2007zza}
  J.~P.~Conlon and D.~Cremades,
  ``The neutrino suppression scale from large volumes,''
  Phys.\ Rev.\ Lett.\  {\bf 99} (2007) 041803
  [arXiv:hep-ph/0611144].

\bibitem{Conlon:2009xf}
  J.~P.~Conlon,
  ``Gauge Threshold Corrections for Local String Models,''
  JHEP {\bf 0904} (2009) 059
  [arXiv:0901.4350 [hep-th]].


\bibitem{Conlon:2009kt}
  J.~P.~Conlon and E.~Palti,
  ``Gauge Threshold Corrections for Local Orientifolds,''
  JHEP {\bf 0909} (2009) 019
  [arXiv:0906.1920 [hep-th]].

\bibitem{Conlon:2009qa}
  J.~P.~Conlon and E.~Palti,
  ``On Gauge Threshold Corrections for Local IIB/F-theory GUTs,''
  arXiv:0907.1362 [hep-th].



\bibitem{Murayama:2001ur}
  H.~Murayama and A.~Pierce,
  ``Not even decoupling can save minimal supersymmetric SU(5),''
  Phys.\ Rev.\  D {\bf 65} (2002) 055009
  [arXiv:hep-ph/0108104].

\bibitem{Raby:2002wc}
  S.~Raby,
  ``Proton decay,''
  arXiv:hep-ph/0211024.

\bibitem{Senjanovic:2009kr}
  G.~Senjanovic,
  ``Proton decay and grand unification,''
  AIP Conf.\ Proc.\  {\bf 1200} (2010) 131
  [arXiv:0912.5375 [hep-ph]].

\bibitem{09052289}
  R.~Tatar, Y.~Tsuchiya and T.~Watari,
  ``Right-handed Neutrinos in F-theory Compactifications,''
  Nucl.\ Phys.\  B {\bf 823} (2009) 1
  [arXiv:0905.2289 [hep-th]].

\bibitem{Leontaris:2009wi}
  G.~K.~Leontaris and N.~D.~Tracas,
  ``Gauge coupling flux thresholds, exotic matter and the unification scale in
  F-SU(5) GUT,''
  arXiv:0912.1557 [hep-ph].


\bibitem{Conlon:2006tj}
  J.~P.~Conlon, D.~Cremades and F.~Quevedo,
  ``Kaehler potentials of chiral matter fields for Calabi-Yau string
  compactifications,''
  JHEP {\bf 0701} (2007) 022
  [arXiv:hep-th/0609180].

\bibitem{Amsler}
C.~Amsler~et~al. (Particle Data Group), Physics Letters B667, 1 (2008)

\bibitem{ArkaniHamed:2000bq}
  N.~Arkani-Hamed, L.~J.~Hall, H.~Murayama, D.~Tucker-Smith and N.~Weiner,
  ``Small neutrino masses from supersymmetry breaking,''
  Phys.\ Rev.\  D {\bf 64} (2001) 115011
  [arXiv:hep-ph/0006312].

\bibitem{Froggatt:1978nt}
  C.~D.~Froggatt and H.~B.~Nielsen,
  ``Hierarchy Of Quark Masses, Cabibbo Angles And CP Violation,''
  Nucl.\ Phys.\  B {\bf 147}, 277 (1979).


\end{thebibliography}
\end{document}